\documentstyle[aps,prb,epsfig,calc]{revtex}
\def\fl{}
\def\etal{\emph{et al.}}
\def\JPCM{\emph{J. Phys.: Condens. Matter} }

\begin{document}
\title{Kinetics of formation of twinned structures under L1$_0$-type
orderings in alloys}
\author{K. D. Belashchenko\cite{Ames}, I. R. Pankratov, 
G. D. Samolyuk\cite{Ames} and V. G. Vaks}
\address{Russian Research Centre `Kurchatov Institute',
Moscow 123182, Russia}

\maketitle

\begin{abstract}
The earlier-developed master equation approach  and kinetic cluster
methods are applied to study kinetics of L1$_0$-type orderings in alloys,
including the formation of twinned structures characteristic of
cubic-tetragonal-type phase transitions. A microscopical model of
interatomic deformational interactions is suggested which generalizes a
similar model of Khachaturyan for dilute alloys to the physically
interesting case of concentrated alloys. The model is used to simulate
A1$\to$L1$_0$ transformations after a quench of an alloy from the
disordered A1 phase to the single-phase L1$_0$ state for a number of
alloy models with different chemical interactions, temperatures,
concentrations, and tetragonal distortions. We find a number of peculiar
features in both transient microstructures and transformation kinetics,
many of  them agreeng well with experimental data. The simulations also
demonstrate a phenomenon of an interaction-dependent alignment of
antiphase boundaries in nearly-equilibrium twinned bands which seems to 
be observed in some experiments.
\end{abstract}

\section{Introduction}\label{intro}

Studies of microstructural evolution under alloy phase transformations
from the disordered FCC phase (A1 phase) to the 
CuAu I-type ordered tetragonal phase (L1$_0$ phase)
attract interest from both fundamental and applied points of view.
A characteristic feature of such transitions is the formation in the 
ordered phase of peculiar `polytwinned' structures
consisting of arrays of
ordered bands separated by the antiphase boundaries 
(APBs) lying in the  (110)-type planes, while the tetragonal axes of 
antiphase-ordered domains (APDs) in the 
 adjacent  bands have `twin-related' (100) and (010)-type orientations
\cite{Leroux,Zhang-91,Zhang-92,Yanar,Tanaka,Oshima,Syutkina}.
Transformation A1$\to$L1$_0$ includes a number of intermediate
stages, including the
`tweed' stage  discussed below. 
These transformations are inherent, in particular, to many alloy systems 
with outstanding magnetic characteristics,
such as  Co--Pt, Fe--Pt, Fe--Pd and similar alloys, and studies of their microstructural
 features, for example, properties and evolution of APBs, 
are interesting for applications of these systems in various magnetic devices 
for which the structure and the distribution of APBs can be very important 
\cite{Zhang-91,Zhang-92,Yanar}. 

The physical reason for the formation of twinned structures was discussed by a number
of authors \cite{Roitburd,Khach-Shat,Khach-book,Vaks-01},
and it is explained by the elimination of the volume-dependent part of elastic 
energy for such structures. However, theoretical treatments of the kinetics of  
A1$\to$L1$_0$ transformation seem to be rather scarce as yet. Khachaturyan and 
coworkers \cite{CWK}
discussed kinetics of tweed and twin formation using a 2D model in a square
lattice with a number of simplifying approximations: a mean-field-type kinetic
equation; a phenomenological description of interaction between elastic strains and
local order parameters; an isotropic elasticity; an unrealistic interatomic interaction
model (with the nearest-neighbour interaction being by an order of magnitude weaker
 than more distant interactions), etc. 
In spite of all these assumptions, some features of evolution
found by Khachaturyan and coworkers \cite{CWK} agree qualitatively with
experimental observations \cite{Zhang-91,Zhang-92,Yanar}. It may illustrate
a low sensitivity of these features to the real structure and 
interactions in an alloy. However, such an oversimplified approach is evidently 
insufficient to study the details of evolution  and their dependence
on the characteristics of an alloy, such as the 
type of interatomic interaction, concentration, temperature, etc,
  which seems to be most  interesting for both applications and 
physical studies of the problem. 

In this work we investigate kinetics of the A1$\to$L1$_0$ transition using
the microscopical  master equation approach and 
the kinetic cluster field method \cite{Vaks-96,BV-98}.
  Earlier this method was used to study
A1$\to$L1$_2$-type transformations \cite{BDPSV} 
as well as early stages of the A1$\to$L1$_0$ transition when 
the deformational interaction $H_d$
 due  to the tetragonal distortion of the L1$_0$ phase is still 
insignificant for the evolution \cite{PV}. Here we consider 
all stages of this transition, including the tweed and
twin stages when the interaction $H_d$ becomes important. 
To this end we first derive a microscopical model
for $H_d$ which generalizes the analogous model of Khachaturyan for dilute alloys 
\cite{Khach-book} to the physically interesting case of concentrated alloys. 
Then we employ the kinetic cluster field method 
to simulate A1$\to$L1$_0$ transformation
in the presence of deformational interaction $H_d$ 
for a number of alloy models with both 
short-range and extended-range chemical interactions at different temperatures, 
concentrations and tetragonal deformations. 
The simulations reveal a number of interesting microstructural features,
 many of them agreeing well with experimental observations
\cite{Zhang-91,Zhang-92,Yanar}. We observe, in particular, 
 a peculiar phenomenon of an interaction-dependent alignment of 
orientations of APBs within twin bands which 
was earlier discussed phenomenologically  \cite{Vaks-01}.
The simulations also show that the type of microstructural evolution 
strongly depends on the interaction type as well as on
the concentration $c$ and temperature $T$.
In particular, drastic, phase-transiton-like  changes in morphology of APBs
within twin bands can occur under variation of $c$ or $T$ in the 
short-range-interaction systems.

The paper is organized as follows. In section 2 we derive
a microscopical expression for the deformational interaction $H_d$ in
concentrated alloys. In section 3 we describe our methods of simulation 
of A1$\to$L1$_0$ transition
which are similar to those used earlier \cite{BDPSV,PV}.
In section 4 we investigate the  transformation kinetics
for the alloy systems with an extended or intermediate
 interaction range, and in section 5, that for the short-range-interaction systems.
Our main conclusions are summarized in section 6.

\section{Model for deformational interaction in concentrated alloys}
\label{Deform-int}

We consider a binary substitutional alloy A$_c$B$_{1-c}$. Various distributions of 
atoms over lattice sites $i$ are described by the sets of
occupation numbers $\{n_i\}$ where the operator $n_{i}=n_{{\rm A}i}$
is unity when the site $i$ is occupied by  atom A and
 zero otherwise. The effective Hamiltonian $H_{\rm eff}$ describing the energy of these
distributions has the form
\begin{equation}
H_{\rm eff}=\sum_{i>j}v_{ij}n_in_j+
\sum_{i>j>k}v_{ijk}n_in_jn_k+\ldots
\label{H_{eff}}
\end{equation}
where $v_{i\ldots j}$ are effective interactions. 

The interactions $v_{i\ldots j}$  include the 
`chemical' contributions $v_{i\ldots j}^c$
which describe the energy changes under the substitution of some atoms A by atoms B
in the rigid lattice, and the `deformational' interactions $v_{i\ldots j}^d$
related to the difference in the lattice deformation under such a substitution.
 A microscopical model for $v^d$ in  dilute alloys 
 was suggested by Khachaturyan \cite{Khach-book}. 
The deformational interaction in concentrated alloys can lead to some new 
effects that are absent in the dilute alloys, in particular, to the
lattice symmetry changes under phase transformations, such as 
the tetragonal distortion under L1$_0$ ordering. Earlier these effects 
 were treated only phenomenologically \cite{CWK}.
Below we describe a microscopical model for calculations of $v^d$
which generalizes the Khachaturyan's approach \cite{Khach-book}
to  the case of concentrated alloys.

Let us denote the position of site $k$ in  the disordered `averaged' crystal 
as ${\bf r}_k$. Because of the randomness of a real disordered or partially
ordered alloy the actual atomic position (averaged over thermal vibrations)
 is not ${\bf r}_k$ but ${\bf r}_k+{\bf u}_k$ where ${\bf u}_k$ is the `static displacement'.
Supposing this displacement  to be small we can expand 
the `adiabatic' (averaged over rapid phonon motion) alloy energy 
$H=H\{n_i,{\bf u}_k\}$ to second order in ${\bf u}_k$:
\begin{equation}
H=H_c\{n_i\}-\sum_k u_{\alpha k}F^{\alpha k}+{1\over 2}\sum_{k,l}
u_{\alpha k}u_{\beta l} A_{\alpha k,\beta l}
\label{H_u}
\end{equation}
where $\alpha$ and $\beta$ are Cartesian indices
and the summation over repeated Greek indices is implied here and below.
The term $H_c\{n_i\}$ 
in (\ref{H_u}) describes interactions in the undistorted average crystal 
lattice, i. e. chemical interactions $v_{i\ldots j}^c$ mentioned above;
$F^{\alpha k}$ can be called the generalized Kanzaki force; and $A_{\alpha k,\beta l}$
is the force constant matrix. Both quantities $F^{\alpha k}$
and $A_{\alpha k,\beta l}$ are certain functions of occupation numbers $n_i$,
and their evaluation needs some further approximations. 

Below we consider ordering phase transitions at the fixed mean concentration $c$. 
Changes of elastic constants and phonon spectra under such transitions
are usually small\cite{Rouchy}.
Therefore, the force constant matrix
$A_{\alpha k,\beta l}$ can be reasonably well approximated with the 
simple `average crystal' approximation:
$A_{\alpha k,\beta l}\{n_i\}\rightarrow
 A_{\alpha k,\beta l}\{c\}\equiv\overline A_{\alpha k,\beta l}$.
To approximate the Kanzaki force $F^{\alpha k}$  we first formally write it as
a series in the occupation numbers $n_i$:
\begin{equation}
F^{\alpha k}\{n_i\}=F^{\alpha k}_0+
\sum_iF^{\alpha k,i}_1n_i+
\sum_{i>j}F^{\alpha k,ij}_2n_in_j+\ldots
\label{F_k}
\end{equation}
Equilibrium values of displacements ${\bf u}_k={\bf u}_k^e\{n_i\}$ 
at the given distribution $\{n_i\}$ are
determined by the minimization of energy (\ref{H_u}) over ${\bf u}_k$,
and the constant $F^{\alpha k}_0$ in (\ref{F_k}) affects only the reference point
${\bf u}_k^e\{0\}$ in the function ${\bf u}_k^e\{n_i\}$.
This constant can be 
determined, for example, from the condition of vanishing of mean static displacements 
in the averaged crystal at some  $c=c_0$,
which implies the relation: $\langle F^{\alpha k}\{n_i\}\rangle_{c=c_0}=0$
where the symbol $\langle \ldots\rangle$  means the statistical averaging
over an alloy. 
The constants $F^{\alpha k}_0$ are insignificant for what follows, and below they
are omitted to simplify formulas. 

In writing an explicit expression for the contribution  $H_K$
(to be called for brevity the `Kanzaki term') of the
occupation-dependent Kanzaki forces
  in energy (\ref{H_u}) one should
consider that due to the translation invariance it can include
only differences of displacements 
$({\bf u}_k-{\bf u}_i)$, 
$({\bf u}_k-{\bf u}_j)$, etc. Therefore, this term should have the form
\begin{equation}
H_K=\sum_{k,i}({\bf u}_i-{\bf u}_k){\bf f}^{k,i}_1n_i+
\sum_{k,ij}({\bf u}_i-{\bf u}_k){\bf f}^{k,ij}_2n_in_j+\ldots
\label{H_K}
\end{equation}
where ${\bf f}^{k,i_1\ldots i_m}_m\equiv {\bf f}^{k}_m$
 are some parameters describing  interaction of lattice deformations with
site occupations.

Representation (\ref{H_K}) for $H_K$ as a sum of
contributions of $m$-site `clusters' proportional to products $n_{i_1}\ldots n_{i_m}$ is
analogous to similar cluster expansions for the `chemical' Hamiltonian
$H_c\{n_i\}$ in (\ref{H_u}). These expansions have been widely discussed, in 
particular, in connection with first-principle calculations of chemical interactions 
\hbox{$v^c_{i_1\ldots i_m}\equiv v^{c,m}$}, see e. g. \cite{Zunger}. The calculations
have shown that the values of $m$-site interactions $v^{c,m}$ in most alloys rapidly
decrease with an increase of $m$, and the pairwise interaction $v^{c,2}$ is usually 
dominant. It is natural to expect that a similar rapid convergence is also
typical for the expansion (\ref{H_K}). Therefore, below we omit 
many-site interactions ${\bf f}^{k}_m$ with $m>2$ in Eq. (\ref{H_K}).
At the same time, 
in estimates of parameters ${\bf f}^{k}_m$ for real alloys below we
combine some model assumptions about  ${\bf f}^{k}_m$ with using of available 
experimental data about the variations of lattice deformations with concentration and 
orderings, and such  estimates may also 
implicitly include the contributions of many-site interactions ${\bf f}^{k}_m$.

For what follows it is convenient to proceed from
functions ${\bf u}_k={\bf u}({\bf r}_k)$,  $n_i=n({\bf r}_i)$,
${\bf f}_1^{k,i}={\bf f}_1({\bf r}_k-{\bf r}_i)$,
${\bf f}_2^{k,ij}={\bf f}_2({\bf r}_k-{\bf r}_i,{\bf r}_j-{\bf r}_i)$ 
and $\overline A_{\alpha k,\beta l}=\overline A_{\alpha\beta}({\bf r}_k-{\bf r}_l)$
in Eqs. (\ref{H_u}) and (\ref{H_K}) to their
Fourier components in the average crystal lattice. Then the energy
(\ref{H_u}) takes the form:
\begin{equation}
H=H_c\{n_i\}+{\frac{1}{N}}\sum_{\bf k}{\bf u}_{-{\bf k}}
\left(n_{\bf k}{\bf f}_{1{\bf k}}+
\sum_{\bf R}\sigma_{\bf k}^{\bf R}{\bf f}_{2{\bf k}}^{\bf R}\right)
+{\frac{1}{2N}}\sum_{\bf k}
u_{-{\bf k}}^{\alpha}\overline A_{\bf k}^{\alpha\beta}u_{{\bf k}}^{\beta}.
\label{H_uk}
\end{equation}
Here $N$ is the total number of crystal cells, the summation over $\bf k$
goes within the Brillouin zone of the averaged crystal,
and we use the following notation:
\begin{eqnarray}
\fl
{\bf u}_{\bf k}=\sum_{\bf r}{\bf u}({\bf r})e^{-i\bf{kr}};\qquad
\hspace{6mm} n_{\bf k}=\sum_{\bf r}n({\bf r})e^{-i\bf{kr}};\qquad
\sigma_{\bf k}^{\bf R}=\sum_{\bf r}n({\bf r})n({\bf r}+{\bf R})e^{-i\bf{kr}};\nonumber\\
\fl
{\bf f}_{1{\bf k}}=\sum_{\bf r}{\bf f}_{1}({\bf r})(1-e^{-i\bf{kr}});\quad
{\bf f}_{2{\bf k}}^{\bf R}=\sum_{\bf r}{\bf f}_{2}({\bf r,\bf R})(1-e^{-i\bf{kr}});\quad
\overline A_{\bf k}^{\alpha\beta}=\sum_{\bf r}\overline A_{\alpha\beta}({\bf r})e^{-i\bf{kr}}.
\label{u_k}
\end{eqnarray}
If one adopts a
commonly used model of `central' Kanzaki forces in which 
forces ${\bf f}_1^{k,i}$ and ${\bf f}_2^{k,ij}$ in (\ref{H_K}) are supposed
to be proportional to the vector ${\bf r}_{ki}=({\bf r}_k-{\bf r}_i)$, 
the vector functions ${\bf f}_{1{\bf k}}$ and ${\bf f}_{2{\bf k}}$ in (\ref{u_k})
can be expressed via two scalar functions, $\varphi_1$ and $\varphi_2$:
\begin{equation}
{\bf f}_{1{\bf k}}=\sum_{\bf r}{\bf r}\,\varphi_1 (r)\,(1-e^{-i\bf{kr}}),\quad
{\bf f}_{2{\bf k}}^{\bf R}=\sum_{\bf r}{\bf r}\,\varphi_2({\bf r,\bf R})\,(1-e^{-i\bf{kr}}).
\label{varphi}
\end{equation}

The functions $\varphi_1$ and $\varphi_2$ in (\ref{varphi}) determine the dependence
of equilibrium lattice parameters on concentration or
ordering. To show it we first note that the homogeneous deformation
$\overline u_{\alpha\beta}$  
is described by Fourier-components ${\bf u}_{\bf k}$ with small 
${\bf k}\to 0$, while 
functions ${\bf f}_{1{\bf k}}$ and ${\bf f}_{2{\bf k}}$ in Eqs. (\ref{H_uk}) and
(\ref{u_k}) at small ${\bf k}$ are linear in ${\bf k}$. Thus the contribution of
homogeneous deformations 
to the Kanzaki term in (\ref{H_uk}) is 
proportional to Fourier-components $u^{\alpha\beta}_{\bf k}$
of the elastic strain $u_{\alpha\beta}=
(\partial u_{\alpha}/\partial x_{\beta}+\partial u_{\beta}/\partial x_{\alpha})/2$
at ${\bf k}\to 0$ and, according to first equation (\ref{u_k}), these
components  are related to
$\overline u_{\alpha\beta}$ as 
\begin{equation}
u^{\alpha\beta}_{\bf k}|_{{\bf k}\to 0}=i(k_{\beta}u_{\bf k}^{\alpha}+
k_{\alpha}u_{\bf k}^{\beta})|_{{\bf k}\to 0}=N\overline u_{\alpha\beta}.
\label{u_strain}
\end{equation}

At small $\bf k$ the force constant matrix $\overline A_{\bf k}^{\alpha\beta}$ 
in (\ref{H_uk}) is bilinear in $\bf k$, and the last term of (\ref{H_uk}) 
corresponds to the standard expression for the elastic energy 
bilinear in $\overline u_{\alpha\beta}$
and linear in the elastic constants 
$c_{\alpha\beta\gamma\delta}$,
see e.g. \cite{Khach-book}. Therefore, the total contribution 
 of terms with  the homogeneous elastic strain $\overline u_{\alpha\beta}$ to energy (\ref{H_uk})
 (to be called `the elastic strain energy' $E_{el}$) can be written as
\begin{equation}
E_{el}=-\overline u_{\alpha\beta}\left(A_1^{\alpha\beta}n_0+
\sum_{\bf R}A_{2{\bf R}}^{\alpha\beta}\sigma_0^{\bf R}\right)+
{\frac{1}{2}}N\Omega\,c_{\alpha\beta\gamma\delta}
\overline u_{\alpha\beta}\overline u_{\gamma\delta}.
\label{E_el0}
\end{equation}
Here $\Omega$ is the volume per atom in the average crystal;
quantities $A_1^{\alpha\beta}$ and $A_{2{\bf R}}^{\alpha\beta}$
are expressed via functions $\varphi_1$ and $\varphi_2$ in (\ref{varphi})
as:
\begin{equation}
A_1^{\alpha\beta}=\sum_{\bf r}x_{\alpha}x_{\beta}\varphi_1({\bf r}), \qquad
A_{2{\bf R}}^{\alpha\beta}=\sum_{\bf r}x_{\alpha}x_{\beta}\varphi_2({\bf r},{\bf R}),
\label{A_1,2}
\end{equation}
where $x_{\alpha}$ is the  Cartesian component of vector ${\bf r}=(x_1,x_2,x_3)$;
and $n_0$ or $\sigma_0^{\bf R}$ is the Fourier component, 
$n_{\bf k}$ or $\sigma_{\bf k}^{\bf R}$, at ${\bf k}=0$.
According to Eq. (\ref{u_k}), the operator $n_0$ or $\sigma_0^{\bf R}$
is the sum of a macroscopically large number $N$ of similar terms.
Thus within the statistical accuracy each of these operators can be substituted 
by its average value:
\begin{equation}
n_0=N\langle n({\bf r})\rangle=Nc; \qquad
\sigma_0^{\bf R}=N\langle n({\bf r})\,n({\bf r}+{\bf R})\rangle.
\label{sigma_0}
\end{equation}
The last average in (\ref{sigma_0}) can be expressed via mean occupations 
of sites and their correlators. In an ordered alloy there exist
several non-equivalent sublattices $s$ with the lattice vectors 
${\bf r}_s$ and mean occupations $c_s=\langle n({\bf r}_s)\rangle$, and so
 the last average in (\ref{sigma_0}) includes averaging over all
sublattices $s$:
\begin{equation}
\langle n({\bf r})\,n({\bf r}+{\bf R})\rangle =\sum_s\nu_s\left(c_s\,c_{s{\bf R}}+
K_{s{\bf R}}\right).
\label{nn_average}
\end{equation}
Here $c_{s{\bf R}}$ is the mean occupation $\langle n({\bf r})\rangle$
for ${\bf r}={\bf r}_s+{\bf R}$;
$\nu_s=N_s/N$ is the relative number of sites in the 
sublattice $s$; and $K_{s{\bf R}}$ is the correlator of occupations of sites 
located at ${\bf r}={\bf r}_s$ and at
${\bf r}={\bf r}_s+{\bf R}$:
\begin{equation}
K_{s{\bf R}}=\langle \left[n({\bf r}_s)-c_s\right]\left[n({\bf r}_s+{\bf R})-c_{s{\bf R}}\right]\rangle
\label{K_R}.
\end{equation}
In a disordered alloy all sites are equivalent,
thus $c_s=c_{s{\bf R}}=c$; \  $\nu_s=1$;  and both index $s$ and the summation
over $s$ in (\ref{nn_average}) are omitted. 

Using Eqs. (\ref{sigma_0}) and (\ref{nn_average}) one can rewrite the elastic strain energy
(\ref{E_el0}) as
\begin{equation}
\hspace{-15mm}
E_{el}=-N\overline u_{\alpha\beta}\left[A_1^{\alpha\beta}\,c+
\sum_{\bf R}\sum_s\nu_s\left(c_s\,c_{s{\bf R}}+
K_{s{\bf R}}\right)A_{2{\bf R}}^{\alpha\beta}\right]+
{\frac{1}{2}}N\Omega\,c_{\alpha\beta\gamma\delta}
\overline u_{\alpha\beta}\overline u_{\gamma\delta}.
\label{E_elK}
\end{equation}
The correlator
$K_{s{\bf R}}$ in Eq. (\ref{E_elK}) can be
calculated using that or another method of statistical theory.
However, for most alloy systems of practical interest, in particular, 
at $c$ and $T$ values not close to the thermodynamic instabilty points 
$T_s$, the correlators $K_{s{\bf R}}$ are 
small and can be neglected. Then equation  (\ref{E_elK}) is simplified:
\begin{equation}
E_{el}=-\overline u_{\alpha\beta}\left[NA_1^{\alpha\beta}\,c+
\sum_{{\bf r},{\bf R}}c({\bf r})c({\bf r}+{\bf R})A_{2{\bf R}}^{\alpha\beta}\right]+
{\frac{1}{2}}N\Omega\,c_{\alpha\beta\gamma\delta}
\overline u_{\alpha\beta}\overline u_{\gamma\delta}.
\label{E_el}
\end{equation}
Equilibrium values of $\overline u_{\alpha\beta}$ in
the absence of applied stress are determined by the minimization of
energy $E_{el}$ with respect to $\overline u_{\alpha\beta}$ which gives:
\begin{equation}
\Omega\,c_{\alpha\beta\gamma\delta}\overline u_{\gamma\delta}=
A_1^{\alpha\beta}\,c+
{\frac{1}{N}}\sum_{{\bf r},{\bf R}}c({\bf r})c({\bf r}+{\bf R})A_{2{\bf R}}^{\alpha\beta}.
\label{u_eq}
\end{equation}
Eq. (\ref{u_eq}) 
enables one to express the equilibrium strain
$\overline u_{\alpha\beta}$ via the concentration, order parameters, and the
interaction parameters $A_1^{\alpha\beta}$ and $A_{2{\bf R}}^{\alpha\beta}$, and it
can also be used to estimate these interaction parameters
from experimental data on  $\overline u_{\alpha\beta}(c,T)$.

Let us consider Eqs. (\ref{E_el}) and (\ref{u_eq}) in particular cases. For a
disordered phase with $c({\bf r})=c$, \  Eq. (\ref{u_eq}) takes the form
\begin{equation}
\Omega\,c_{\alpha\beta\gamma\delta}\overline u_{\gamma\delta}=
A_1^{\alpha\beta}\,c+A_2^{\alpha\beta}\,c^2
\label{A_12c}
\end{equation}
where $A_2^{\alpha\beta}=\sum_{\bf R}A_{2{\bf R}}^{\alpha\beta}$. If the 
disordered phase has a
cubic symmetry (as for the FCC or BCC alloys), quantities 
$A_1^{\alpha\beta}$ and $A_2^{\alpha\beta}$ are proportional to the Kronecker symbol 
$\delta_{\alpha\beta}$, and Eq. (\ref{A_12c}) determines the concentrational dilatation 
$u(c)=\overline u_{\alpha\alpha}(c)-u_{\alpha\alpha}(0)$:
\begin{equation}
u(c)=(A_1c+A_2c^2)/\Omega B.
\label{u_c}
\end{equation}
Here $B=(c_{11}+2c_{12})/3$ is the bulk modulus;
$c_{ij}$ are the elastic constants 
in  Voigt's notation; and 
coefficients $A_1$ and $A_2$ are expressed via functions
$\varphi_1$ and $\varphi_2$ in (\ref{varphi}), (\ref{A_1,2}):
\begin{equation}
A_1=\sum_{\bf r}\varphi_1({\bf r})r^2/3;\qquad 
A_2=\sum_{{\bf r},{\bf R}}\varphi_2({\bf r},{\bf R})r^2/3.
\label{A_1}
\end{equation}
The linear in $c$  term  in (\ref{u_c}) corresponds to the Vegard law
while the term with $A_2$ describes the  non-linear deviations from 
this law. Such deviations were observed for many alloys, and
these data can be used to estimate $A_2$ values, but in these 
estimates one should also take into consideration a possible concentration
dependence of the bulk modulus $B$.

For the ordered phase, the mean occupation $c({\bf r})$
 can be written as a superposition of concentration waves
corresponding to certain superstructure vectors ${\bf k}_p$ \cite{Khach-book}:
\begin{equation}
c({\bf r})=c+{1\over 2}\sum_p\left[\eta_p\exp(i{\bf k}_p{\bf r})+
\eta_p^*\exp(-i{\bf k}_p{\bf r})\right],
\label{cr_CW}
\end{equation}
and amplitudes $\eta_p$ can be considered  as order parameters.
After the substitution of expressions (\ref{cr_CW}) for $c({\bf r})$ and 
$c({\bf r}+{\bf R})$ in Eq. (\ref{E_el}) the linear in $\eta_p$ terms 
vanish due to the crystal symmetry, and the first term
of (\ref{E_el}) becomes the sum of 
the ordering-independent term and the term bilinear in order parameters:
\begin{equation}
E_{el}=-N\overline u_{\alpha\beta}\left(A_1^{\alpha\beta}\,c+
A_2^{\alpha\beta}\,c^2+\sum_pq_{\alpha\beta pp}|\eta_p|^2\right)+
{\frac{1}{2}}N\Omega\,c_{\alpha\beta\gamma\delta}
\overline u_{\alpha\beta}\overline u_{\gamma\delta}.
\label{E_el-q}
\end{equation}
Here quantities $q_{\alpha\beta pp}$ have a  different form
in the cases ($a$) when the superstructure vector ${\bf k}_p$ is half
of some reciprocal lattice vector ${\bf g}$ and thus both the
order parameter $\eta_p$ and all factors  $\exp (i{\bf k}_p{\bf r})$ in (\ref{cr_CW})
are real, and ($b$) when ${\bf k}_p\neq {\bf g}/2$:
\begin{equation}
(a)\qquad {\bf k}_p={\bf g}/2:\hspace{10mm}
q_{\alpha\beta pp}=\sum_{{\bf r},{\bf R}}x_{\alpha}x_{\beta}
\varphi_2({\bf r},{\bf R})\exp\, (i{\bf k}_p{\bf r});
\label{q_a}
\end{equation}
\begin{equation}
(b)\qquad {\bf k}_p\neq {\bf g}/2:\hspace{10mm}
q_{\alpha\beta pp}={1\over 2}
\sum_{{\bf r},{\bf R}}x_{\alpha}x_{\beta}
\varphi_2({\bf r},{\bf R})\cos\, ({\bf k}_p{\bf r}).
\label{q_b}
\end{equation}
The coefficients $q_{\alpha\beta pp}$ in (\ref{E_el-q}) (to be called 
the `striction' coefficients, in an analogy with the terminology used 
in the ferroelectricity or magnetism theory) are commonly used in  
phenomenological theories of lattice distortions
under orderings \cite{Roitburd,Khach-Shat,Khach-book,Vaks-01,CWK}.
Eqs. (\ref{q_a}), (\ref{q_b}) and (\ref{A_1,2}) provide the microscopic expression for 
these coefficients via the
function  $\varphi_2$ describing
non-pairwise Kanzaki forces in Eqs.  (\ref{H_uk})--(\ref{varphi}).

Let us apply Eqs. (\ref{cr_CW})--(\ref{q_a}) to the case of 
L1$_0$ or L1$_2$ ordering in FCC alloys which are described by
three real order parameters $\eta_{\alpha}$ \cite{Khach-book,BDPSV}.
Eqs. (\ref{cr_CW}) here take the form
\begin{equation}
c({\bf r})=c+\eta_1\exp (i{\bf k}_1{\bf r})+\eta_2\exp (i{\bf k}_2{\bf r})
+\eta_3\exp (i{\bf k}_3{\bf r}),
\label{c_r}
\end{equation}
where  ${\bf k}_{\alpha}={\bf g}_{\alpha}/2$ is the superstructure vector corresponding
to $\eta_{\alpha}$:
\begin{equation}
{\bf k_1}=[100]2\pi/a,\qquad {\bf k_2}=[010]2\pi/a,\qquad{\bf k_3}=[001]2\pi/a.
\label{k_alpha}
\end{equation}
In the cubic L1$_2$  structure one has: $|\eta_1|=|\eta_2|=|\eta_3|$, $\eta_1\eta_2\eta_3>0$,
and four types of ordered domains are possible. In the L1$_0$-ordered structure 
with the tetragonal axis $\alpha$ a single
parameter $\eta_{\alpha}$ is present which is either positive or negative, 
and so  six types of ordered domains 
are possible.

The striction coefficients 
for L1$_0$ or L1$_2$ ordering are determined by Eq. 
(\ref{q_a}). Due to the cubic symmetry of the `average' FCC crystal,
there are only two different striction coefficients,
$q_{1111}$ and $q_{1122}$ (and those obtained from them by the cubic
symmetry operations), which for brevity will  be denoted as $q_{11}$ and $q_{12}$,
respectively:
\begin{equation}
q_{11}=\sum_{{\bf r},{\bf R}}x_1^2\varphi_2({\bf r},{\bf R})\exp (i{\bf k}_1{\bf R});\quad
q_{12}=\sum_{{\bf r},{\bf R}}x_1^2\varphi_2({\bf r},{\bf R})\exp (i{\bf k}_2{\bf R}).
\label{q_12}
\end{equation}
Variation of elastic constants 
$c_{\alpha\beta\gamma\delta}$ with ordering is usually small\cite{Rouchy}, and for simplicity it
will be neglected. Then minimizing energy (\ref{E_el-q})
with respect to $\overline u_{\alpha\beta}$ we obtain the expressions for 
lattice deformations induced by ordering  (\ref{c_r}):
\begin{equation}
%\fl
\hspace{-20mm}\overline u =q_+(\eta_1^2+\eta_2^2+\eta_3^2)/\Omega c_+,\quad
\overline\varepsilon  =q_-[\eta_1^2-(\eta_2^2+\eta_3^2)/2]/\Omega c_-,\quad
\zeta =q_-(\eta_2^2-\eta_3^2)/\Omega c_-.
\label{overline_u}
\end{equation}
Here  $\overline u=\overline u_{11}+\overline u_{22}+\overline u_{33}$
describes the volume change; 
$\overline\varepsilon  =\overline u_{11}-(\overline u_{22}+\overline u_{33})/2$
is the tetragonal distortion; 
$\zeta =\overline u_{22}-\overline u_{33}$ is the shear deformation; and
$q_{\pm}$ or $c_{\pm}$ are  linear combinations of striction or elastic constants:
\begin{equation}
%\fl
\hspace{-10mm} q_-=q_{11}-q_{12}; \quad  c_-=c_{11}-c_{12};\quad q_+=q_{11}+2q_{12};  \quad
c_+=c_{11}+2c_{12}.
\label{q_pm}
\end{equation}
For the  L1$_2$ ordering, values $|\eta_1|=|\eta_2|=|\eta_3|=\eta $ are the same, 
so just the volume striction $\overline u=3q_+\eta^2/\Omega c_+$ is present, while
in the L1$_0$-ordered domain with  $\eta_2=\eta_3=0$ one has  both the
volume  and the tetragonal striction:
\begin{equation}
\overline u=q_+\eta_1^2/\Omega c_+,\qquad 
\overline\varepsilon  =q_-\eta_1^2/\Omega c_-.
\label{overline_u-eta1}
\end{equation}
 Therefore, using experimental data about the lattice distortions
and order parameters  under L1$_2$ and L1$_0$ orderings 
 one can estimate the striction 
coefficients $q_{11}$ and $q_{12}$ 
and thus the non-pairwise Kanzaki interaction $\varphi_2$ in Eqs. (\ref{q_12}).

Below we suppose for simplicity the interaction
$\varphi_2({\bf r},{\bf R})$ to be short-ranged, i. e.
significant only when each of three relative distances
$r$, $R$ and $|{\bf r}-{\bf R}|$ does not exceed the nearest-neighbour distance
$\rho =a/\sqrt{2}$. Then this function  can be written as
\begin{equation}
\varphi_2({\bf r},{\bf R})=\delta_{r,\rho}\delta_{R,\rho}\left(\varphi_{a}\,\delta_{|{\bf r}-{\bf R}|,0}
+\varphi_{b}\,\delta_{|{\bf r}-{\bf R}|,\rho}\right)
\label{varphi_ab}
\end{equation}
where $\delta_{r,\rho}$ is the Kronecker symbol equal to unity when 
$r=\rho$ and zero otherwise while  $\varphi_{a}$ and $\varphi_{b}$ are 
the interaction parameters.
 The assumption (\ref{varphi_ab}) is analogous
to that used by Khachaturyan \cite{Khach-book} for the pairwise
Kanzaki interaction $\varphi_1({\bf r})$ in (\ref{varphi}):
\begin{equation}
\varphi_1({\bf r})=\varphi_1\delta_{r,\rho}
\label{varphi_1}
\end{equation}
where the constant $\varphi_1$ is estimated from experimental data on
concentrational dilatation.
First-principle estimates of lattice distortions in dilute alloys
\cite{BSVZ} seem to imply that the assumption (\ref{varphi_1}) yields the correct 
order of magnitude of $\varphi_1({\bf r})$. Therefore, the analogous assumption 
(\ref{varphi_ab})
for $\varphi_2({\bf r},{\bf R})$ can be reasonable, too. 

Substituting Eq. (\ref{varphi_ab}) into (\ref{q_12}) we obtain the explicit expression
for coefficients $q_{ik}$  via parameters
 $\varphi_{a}$ and $\varphi_{b}$ in (\ref{varphi_ab}):
\begin{equation}
q_{11}=-2a^2\varphi_{a};\qquad q_{12}=-4a^2\varphi_{b}.
\label{q_varphi}
\end{equation}

 The coefficient  ${\bf f}_{1{\bf k}}$
in Eqs. (\ref{H_uk}) and (\ref{u_k})  for model (\ref{varphi_1}) has the form \cite{Khach-book}:
\begin{equation}
{\bf f}_{1{\bf k}}=4\varphi_1i\sum_{\alpha =1}^3{\bf a}_{\alpha}\sin\,({\bf ka}_{\alpha})
\sum_{\beta\neq\alpha}\cos\,({\bf ka}_{\beta})
\label{f_1k}
\end{equation}
where ${\bf a}_{\alpha}$  is ${\bf e}_{\alpha}a/2$ and ${\bf e}_{\alpha}$ is the unit vector 
along the main crystal axis $\alpha$. The function ${\bf f}_{2{\bf k}}^{\bf R}$
in (\ref{H_uk}), (\ref{u_k})
for model (\ref{varphi_ab}) is the sum of two terms:
\begin{equation}
{\bf f}_{2{\bf k}}^{\bf R}={\bf f}_{a{\bf k}}^{\bf R}+{\bf f}_{b{\bf k}}^{\bf R}.
\label{f_2k}
\end{equation}
Here, ${\bf f}_{a{\bf k}}^{\bf R}$ is $\varphi_a\delta_{R,\rho}{\bf R}\left(1-e^{-i{\bf kR}}\right)$,
while the function  ${\bf f}_{b{\bf k}}^{\bf R}$ for ${\bf R}$ equal to 
${\bf R}_{\alpha}+ {\bf R}_{\beta}$
(where ${\bf R}_{\alpha}$ is ${\bf a}_{\alpha}$ or $(-{\bf a}_{\alpha})$,
${\bf R}_{\beta}$ is ${\bf a}_{\beta}$ or $(-{\bf a}_{\beta})$, and $\beta\neq\alpha$)
 can be written as
\begin{equation}
\fl
{\bf f}_{b{\bf k}}^{\bf R}=2\varphi_b%\delta_{R,\rho}
\left[{\bf R}+\left(i{\bf a'}\sin {\bf ka'}-{\bf R}_{\alpha}\cos\,{\bf ka'}\right)e^{-i{\bf kR}_{\alpha}}
+\left(i{\bf a'}\sin {\bf ka'}-{\bf R}_{\beta}\cos\,{\bf ka'}\right)e^{-i{\bf kR}_{\beta}}\right]
\label{f_bk}
\end{equation}
where ${\bf a'}$ is $[{\bf e}_{\alpha}{\bf e}_{\beta}]a/2$.

Relations (\ref{varphi}), (\ref{varphi_ab})--(\ref{f_bk}) together with 
(\ref{u_c}) and (\ref{q_varphi}) provide a simplified model for the Kanzaki term
$H_K$ in Eqs. (\ref{H_K}) and (\ref{H_uk}). This model 
will be used  below in simulations of  A1$\to$L1$_0$ transitions.
To get an idea about the actual scale of parameters of this model,
let us estimate quantities $q_{ik}$, $\varphi_a$ and $\varphi_b$ in 
Eqs. (\ref{varphi_ab}) and (\ref{q_varphi}) for
the alloys Co--Pt for which detailed data about the lattice distortion under 
L1$_2$ and L1$_0$ orderings are available \cite{Berg-Cohen,Leroux-88}.
The volume change $\overline u$ under both the
L1$_2$ ordering in CoPt$_3$ and L1$_0$ ordering in CoPt appears to be very small
\cite{Berg-Cohen,Leroux-88,Leroux}: $\overline u\lesssim 10^{-3}$. According
to Eqs. (\ref{overline_u}) and (\ref{overline_u-eta1}) it implies the 
relation: $q_{12}\simeq q_{11}/2$.
The value  $q_-=(q_{11}-q_{12})$ for CoPt can be estimated 
from second equation (\ref{overline_u-eta1}) using data of Ref.
\cite{Leroux-88} 
 for $\eta_1$ and $\overline\varepsilon $ 
at $T=0.84\,T_c$:
 \ $\eta_1\simeq 0.4$;
\  $\varepsilon\simeq -0.04$ (with the thermal expansion effect subtracted);
and for the atomic volume:  $\Omega =\Omega (T_c+0)\simeq 13.8{\rm\AA}^3$.
Using also for the elastic constant $c_-=(c_{11}-c_{12})$ its value for the 
FCC platinum, $c_-\simeq 0.97$ Mbar  \cite{Ducastelle}, 
we obtain: $q_-\simeq 2.6\cdot 10^4$ K. Combining it
with the above-mentioned relation $q_{12}\simeq q_{11}/2$ and using Eq. (\ref{overline_u})
 we find: $\varphi_a\simeq 2\cdot 10^4$ K$/a^2$, and
$\varphi_b\simeq 5\cdot 10^3$ K$/a^2$. Let us also note that the 
ordering-induced  elastic energy per atom  $\varepsilon _{el}^{ord}$
 in the CoPt alloy is  small:
$\varepsilon _{el}^{ord}\simeq
\Omega c_-\overline\varepsilon ^2/6\simeq$ \hbox{30 K}, which 
 is much less than the L1$_0$ ordering temperature $T_c\simeq 1100$ K.

The equilibrium values of displacements ${\bf u}_{\bf k}^e={\bf u}_{\bf k}^e(n_i)$ are found by 
the minimization of energy (\ref{H_uk}) over ${\bf u_k}$.
Substituting these ${\bf u}_{\bf k}^e$ into Eq. (\ref{H_uk}) we obtain  the effective
Hamiltonian $H=H_c+H_d$ where the deformational interaction
$H_d$ can be written as
\begin{equation}
\fl
H_d=-{\frac{1}{2N}}\sum_{\bf k}
\left(n_{-\bf k}{\bf f}_{1{\bf k}}^*+
\sum_{\bf R}\sigma_{-\bf k}^{\bf R}{\bf f}_{2{\bf k}}^{{\bf R}*}\right)
{\bf G_k}
\left(n_{\bf k}{\bf f}_{1{\bf k}}+
\sum_{\bf R}\sigma_{\bf k}^{\bf R}{\bf f}_{2{\bf k}}^{\bf R}\right)=H_{d2}+H_{d3}+H_{d4}.
\label{H_d}
\end{equation}
Here the matrix ${\bf G_k}=G_{\bf k}^{\alpha\beta}$ is  inverse to the force constant matrix
$\overline A_{\bf k}^{\alpha\beta}$, and the matrix product $\bf aBc$ means the sum 
$a_{\alpha}B_{\alpha\beta}b_{\beta}$. The term $H_{d2}$,
$H_{d3}$ and $H_{d4}$ in (\ref{H_d}) describes the pairwise, three-particle and
four-particle deformational interaction, respectively:
\begin{equation}
\fl
H_{d2}={\frac{1}{2}}\sum_{\bf r, r'}
n({\bf r})\Phi_2({\bf r}-{\bf r'})n({\bf r'}),\qquad H_{d3}=
{\frac{1}{2}}\sum_{\bf r, r'}\sum_{\bf R}
n({\bf r})\Phi_3^{\bf R}({\bf r}-{\bf r'})n({\bf r'})n({\bf r'}+{\bf R}),
\label{H_d2-3}
\end{equation}
\begin{equation}
H_{d4}={\frac{1}{2}}\sum_{\bf r, r'}\sum_{\bf ,R,R'}n({\bf r})n({\bf r}+{\bf R})
\Phi_4^{\bf R,R'}({\bf r}-{\bf r'})n({\bf r'})n({\bf r'}+{\bf R'}),
\label{H_d4}
\end{equation}
where the potential
$\Phi_2$, $\Phi_3^{\bf R}$ or
$\Phi_4^{\bf R,R'}$ is given by the expression:
\begin{equation}
\fl
\left\{\Phi_2({\bf r});  \Phi_3^{\bf R}({\bf r}); 
\Phi_4^{\bf R,R'}({\bf r})\right\}
=-{\frac{1}{N}}\sum_{\bf k}e^{i{\bf kr}}\left\{{\bf f}_{1{\bf k}}^*{\bf G_k}{\bf f}_{1{\bf k}};\ 
{\bf f}_{1{\bf k}}^*{\bf G_k}{\bf f}_{2{\bf k}}^{{\bf R}}+
{\bf f}_{2{\bf k}}^{{\bf R}*}{\bf G_k}{\bf f}_{1{\bf k}};\ 
{\bf f}_{2{\bf k}}^{{\bf R}*}{\bf G_k}{\bf f}_{2{\bf k}}^{{\bf R'}}\right\}.
\label{Phi_234}
\end{equation}

As the matrix ${\bf G_k}$ in (\ref{Phi_234}) at small $k$ includes the
well-known `elastic singularity' \cite{Khach-book}: ${\bf G_k}\sim 1/k^2$,
each of terms $H_{d2}$, $H_{d3}$ and $H_{d4}$ in (\ref{H_d})
 includes the long-ranged elastic interaction. The formation of twinned structures
discussed below is determined by the four-particle interaction $H_{d4}$.
The rest deformational interactions, $H_{d2}$ and $H_{d3}$,
for the single-phase L1$_0$ ordering under consideration lead just to some 
quantitative renormalizations of chemical interaction $H_c$ in (\ref{H_uk})
which are  usually small and insignificant. Therefore, below we retain in 
the deformational interaction (\ref{H_d}) only the last term $H_{d4}$. 
Let us also note that each term in  the sum 
(\ref{H_d4}) for $H_{d4}$ at fixed $\bf r$ and $\bf r'$ has the order of magnitude
of  the above-mentioned ordering-induced elastic energy $\varepsilon_{el}^{ord}$
which usually is small.
Thus the interaction $H_{d4}$ can be  significant  only because of  `coherent'
contributions of many sites $\bf r$ and $\bf r'$ due to the long-ranged
elastic interaction. Therefore, local fluctuations of occupations
$n({\bf r})$ in the interaction  $H_{d4}$ are insignificant, and it
 can be treated in
the `kinetic mean-field approximation' (KMFA) \cite{Vaks-96,BV-98,BDPSV}
which neglects such fluctuations and corresponds to the 
substitution  in (\ref{H_d4}) of each occupation
operator $n({\bf r})$ by its mean value
$c({\bf r})=\langle n({\bf r})\rangle$ where $\langle \ldots\rangle$ means averaging 
 over the space-  and time-dependent distribution function \cite{Vaks-96,BV-98,BDPSV}. 
Therefore, in considerations of A1$\to$L1$_0$ transformations below we 
approximate the total effective Hamiltonian $H$ in (\ref{H_uk}) by the following
expression:
\begin{equation}
\hspace{-20mm}H=H_c+H_{d4}=H_c\{n({\bf r})\}+
{\frac{1}{2}}\sum_{\bf r, r',R,R'}c({\bf r})c({\bf r}+{\bf R})
\Phi_4^{\bf R,R'}({\bf r}-{\bf r'})c({\bf r'})c({\bf r'}+{\bf R'})
\label{H_cd}
\end{equation}
where the potential $\Phi_4^{\bf R,R'}({\bf r})$ is given by the last equation (\ref{Phi_234}).
 
\section{Models and methods of simulation}

To simulate A1$\to$L1$_0$ transformations in an alloy with the Hamiltonian  (\ref{H_cd})
 we use the methods 
described in Refs. \cite{BDPSV} and \cite{PV} to be referred 
to as I and II, respectively. Evolution of atomic distributions
is described by the 
kinetic tetrahedron cluster field method \cite{BDPSV} in which mean occupations
$c_i=c({\bf r}_i)=\langle n({\bf r}_i)\rangle$ averaged over the space-
and time-dependent distribution function obey the kinetic equation (I.10):
\begin{equation}
dc_i/dt=2\sum_jM_{ij}
\sinh [\beta (\lambda_j-\lambda_i)/2].
\label{c_i(t)}
\end{equation}
Here $\beta =1/T$ is the inverse temperature;
 $M_{ij}$ is the generalized mobility proportional
to the configurationally independent factor  $\gamma_{nn}$ in the
probability of an inter-site atomic exchange  A$i\leftrightarrow {\rm B}j$ between 
neighbouring sites $i$ and $j$ per unit time; and
$\lambda_i=\lambda_i\{c_j\}$ is the local chemical potential
equal to the derivative of the generalized free energy 
$F\{c_i\}$ defined in Refs. \cite{Vaks-96,BV-98} with respect to $c_i$:
$\lambda_i=\partial F/\partial c_i$.  
The expression for 
$M_{ij}=M_{ij}\{c_k\}$  employed in our simulations is given by Eq. (I.12) with the
asymmetrical potential $u_i$ taken zero for simplicity,
while the local chemical potential $\lambda_i$ now is the sum of the chemical
and the deformational term, $\lambda_i^c$ and $\lambda_i^d$. The microscopical 
expressions for $\lambda_i^c$ 
are given by equations (I.13) -- (I.16) which include only chemical interactions
$v_{ij}=v_{ij}^c$, while the deformational contribution $\lambda_i^d=
\lambda_i^d({\bf r}_i)$ is the variational derivative of the second term in
(\ref{H_cd}) with respect to $c_i=c({\bf r}_i)$:
\begin{equation}
\hspace{-5mm} \lambda_i^d({\bf r})=\delta H_{d4}/\delta c({\bf r})=
2\sum_{\bf r',R,R'}c({\bf r}+{\bf R})
\Phi_4^{\bf R,R'}({\bf r}-{\bf r'})c({\bf r'})c({\bf r'}+{\bf R'}).
\label{lambda_d}
\end{equation}

For the chemical interaction $v_{ij}^c$ we 
employ the  five alloy models used  in  I and II:

\ 

1. The second-neighbour interaction model with the nearest-neighbour
interaction $v_1=1000$\,K (in the Boltzmann constant $k_B$ units)
and $v_2/v_1=\epsilon =-0.125$.

2. The same model with $\epsilon =-0.25$.

3. The  same model with $\epsilon =-0.5$.

4. The fourth-neighbour interaction model with $v_n$ estimated by Chassagne \etal 
\cite{Chassagne} from their experimental data for disordered Ni--Al alloys:
 $v_1=1680\,\mathrm{K}$, $v_2=-210\,\mathrm{K}$,
$v_3=35\,\mathrm{K}$, and $v_4=-207\,\mathrm{K}$.

5. The fourth-neighbour interaction model with $v_1=1000$\,K, 
$v_2/v_1=-0.5$, \  $v_3/v_1=0.25$, and  \ $v_4/v_1=-0.125$. 

\ 

The effective interaction range $R_{int}$ for these models monotonously increases with the
model number. Therefore, a comparison of the simulation results for these models
enables one to study the influence of $R_{int}$  on the microstructural
evolution. The critical temperature $T_c$ for the phase 
transition A1$\to$L1$_0$ in the absence of deformational interaction $H_{d4}$
(which seems to have little effect on $T_c$ in our simulations)
 for model 1, 2, 3, 4 and 5 is
614, 840, 1290, 1950 and 2280 K, respectively \cite{VS}.

For the Kanzaki force 
${\bf f}_{2{\bf k}}^{\bf R}$ entering the expression
(\ref{Phi_234}) for the potential $\Phi_4^{\bf R,R'}({\bf r})$ 
 in  (\ref{lambda_d})  we use Eqs.
(\ref{f_2k}) and (\ref{f_bk}). The interaction parameters $\varphi_a$ and $\varphi_b$
in these equations can be expressed via spontaneous deformations 
$\overline u$ and $\overline\varepsilon$ using Eqs. 
(\ref{overline_u-eta1}) and (\ref{q_varphi}).
For simplicity we assume the volume striction to be small
(as it is for the Co--Pt alloys mentioned above):  $\overline u\simeq 0$,
while the tetragonal distortion will be characterized by its maximum value $\varepsilon_m$
in a stoichiometric alloy, i. e. by the value $\overline\varepsilon$ 
in (\ref{overline_u-eta1}) at $\eta_1=0.5$.
Therefore,  interactions $\varphi_a$ and $\varphi_b$
in our simulations are determined by the relations:
\begin{equation}
\varphi_a=-a(c_{11}-c_{12})\varepsilon_m /3;\qquad
\varphi_b=a(c_{11}-c_{12})\varepsilon_m /12.
\label{varphi_epsilon}
\end{equation}
%
%and $\varepsilon_m$ is considered as a simulation parameter. 
For the lattice constant $a$ in (\ref{varphi_epsilon}) we take a typical value
$a\simeq 4$ \AA, and for the elastic constant $(c_{11}-c_{12})$, the value $0.97$ Mbar
corresponding to FCC platinum \cite{Ducastelle}.
%The maximum distortion $\varepsilon_m$ values were varied
%from 0.06 to 0.2, and the influence of  these variations 
% on evolution  was studied in detail.
%and was found to be reduced mainly to some scaling of characteristic
%lengths in the microstructures.

For the force constant matrix $\overline {\bf A}_{\bf k}$ (which determines the
matrix ${\bf G_k}=(\overline {\bf A}_{\bf k})^{-1}$ in Eq. (\ref{Phi_234}))
we use the model described in Refs. \cite{BSV,BDPSV}. It corresponds to a
 Born-von Karman model 
with the first- and second-neighbour force constants only, 
and the second-neighbour constants are supposed to correspond
to a spherically symmetrical interaction. This model
includes three independent force constants which are expressed in terms
of elastic constants $c_{ik}$, and these constants were chosen equal to those 
of the FCC platinum \cite{Ducastelle}: 
$c_{11}=3.47$ Mbar, $c_{12}=2.5$ Mbar, and 
$c_{44}=0.77$ Mbar.

As it was discussed in I and II, the transient partially ordered alloy states
can be described using either mean occupations 
$c_i=c({\bf r}_i)$ or
local order parameters  $\eta_{\alpha i}^2$ and 
local concentrations $\overline c_i$ 
defined by Eqs. (I.24) and (I.25). The simulation results below are
 usually presented as the distributions of quantities 
$\eta_i^2=\eta_{1i}^2+\eta_{2i}^2+\eta_{3i}^2$, to be called the `$\eta^2$--representation', and these
distributions are similar to those observed in the experimental transmission 
electron microscopy (TEM) images \cite{BDPSV}.

Our simulations were performed in the FCC simulation boxes of sizes
$V_b=L^2\times H$ (where $L$ and $H$ are given in the lattice constant $a$ units)
with periodic boundary conditions. We used both quasi-2D simulations
with $H=1$ and  3D simulation with $H\sim L$.
For the given coordinate $z=na$ (with $n=0$ for 2D simulation) each of figures
below shows all FCC lattice 
sites lying in two adjacent planes, $z=na$ and $z=(n+1/2)a$.
The point  $(x,y)$ with  $(x/a, y/a)$ equal to $(l,m)$, $(l+1/2,m)$, $(l+1/2,m+1/2)$
or $(l,m+1/2)$ in the figures corresponds to the lattice site with 
 $(x/a, y/a, z/a)$ equal to $(l,m,n)$, $(l+1/2,m,n+1/2)$, $(l+1/2,m+1/2,n)$
or $(l,m+1/2,n+1/2)$, respectively. Therefore, at $V_b=L^2\times H$ the figure
shows $4L^2$  lattice sites.

The simulation methods were the same as in I and II. 
In simulations of A1$\to$L1$_0$ transformation
 the initial as-quenched
distribution $c_i(0)$ was characterized by its mean value $c$ and small
random fluctuations $\delta c_i$; usually we used $\delta c_i=\pm 0.01$.
 The distribution of initial fluctuations $\delta c_i$ for the given 
simulation box volume $V_b$  was identical  for all models 
and the same as that used in II. 
The sensitivity of simulation results to variations of these initial
fluctuations $\delta c_i$ was discussed in II and was found to be insignificant
for the features of evolution discussed below.

\section{Kinetics of A1$\to$L1$_0$ transformations in systems 
with an extended or intermediate interaction range}

As discussed in I, II and below, features of microstructural evolution 
under A1$\to$L1$_0$ and A1$\to$L1$_2$ transitions
sharply depend on the effective interaction range $R_{int}$ in an alloy.
In this section we discuss A1$\to$L1$_0$ transitions for
the systems with an extended or intermediate interaction range, 
such as our models 5 and 4,
while the short-range-interaction systems are considered in the next section.
  
Some results of our simulations are presented in figures~1--8.
The symbol A or
$\overline {\rm A}$
in these figures corresponds to
an L1$_0$-ordered domain with the tetragonal axis $c$ along (100) and
the positive or negative value, respectively, of  the order parameter  $\eta_{1}$; 
the symbol B or
$\overline {\rm B}$, to that for the $c$-axis along (010) and the order parameter  $\eta_{2}$;
 and the symbol C or $\overline {\rm C}$, to that  for the $c$-axis 
along  (001) and the order parameter  $\eta_{3}$. Figure \ref{m4s-3D} shown in 
the $c$-representation illustrates the occupation of lattice sites for each domain type.
The APB separating two APDs with the same  tetragonal axis 
(i. e. APDs A and $\overline {\rm A}$, B and $\overline {\rm B}$ or C 
and $\overline {\rm C}$) will 
be for brevity called  the `shift-APB',
and the  APB separating the APDs with 
perpendicular tetragonal axes will be called the `flip-APB'.

Before discussing  figures 1--8 we remind the general ideas
about the formation of twinned structures [1--11].
%\cite{Leroux,Zhang-91,Zhang-92,Yanar,Tanaka,Oshima,Syutkina,Roitburd,Khach-Shat,Khach-book,Vaks-01}.
 To avoid discussing the problems of nucleation, in this work we consider
the transformation temperatures $T$ 
lower than the ordering spinodal temperature $T_s$.
Then the evolution under A1$\to$L1$_0$  transition includes
the following stages \cite{Zhang-91,Zhang-92,Yanar,Tanaka,Oshima}:

(i) The initial stage of the formation of  finest  L1$_0$-ordered domains
when  their tetragonal distortion makes still little effect on the evolution
and all six types of APD are present in microstructures in the same proportion.
It corresponds to  the
so-called `mottled' contrast in TEM images
\cite{Tanaka,Oshima}.

(ii) The next, intermediate stage which corresponds to the so-called `tweed' contrast in 
 TEM images. The tetragonal deformation of the  
L1$_0$-ordered APDs here  leads to the predominance of the (110)-type orientations of
flip-APBs, but all six types of APD (i. e. APDs with all three orientations of the 
tetragonal axis $c$) are still present  in microstructures in  comparable 
proportions \cite{Zhang-91,Zhang-92,Yanar}.

(iii) The final, polytwinned stage when the tetragonal distortion
of the L1$_0$-ordered APDs  becomes
the main factor of  evolution and leads to the
formation of (110)-type oriented twin bands.
 Each band includes only two types of APD with the same 
$c$ axis, and  these axes in the adjacent bands  are `twin' related, \hbox{i. e.}  have the 
alternate (100) and (010) orientations for the given set of the (110)-oriented bands 
\cite{Zhang-91,Zhang-92,Yanar}.

The thermodynamic driving force for the (110)-type orientation of flip-APBs
is the gain in the elastic energy of adjacent APDs: at other orientations
 this  energy  increases under the growth of an APD
proportionally to its volume \cite{Roitburd,Khach-Shat,Khach-book,Vaks-01}.
For an APD with the characteristic size $l$ and the surface $S_d$,
 this elastic energy $E_{el}^v\sim c_-\overline\varepsilon^2S_dl$ 
begins to affect the microstructural  evolution when it
becomes comparable with the surface energy
$E_s\sim\sigma S_d$ where $\sigma$ is the APB surface tension.
 The `tweed' stage (ii) corresponds to the relation
$E_{el}^v\sim E_s$ or to the characteristic 
 APD size  
\begin{equation}
l_0\sim \sigma/c_-\overline\varepsilon^2,
\label{l_0}
\end{equation}
and so this size sharply increases under decreasing
distortion  $\overline\varepsilon$.

Figures 1--7 illustrate quasi-2D simulations for which microstructures include
only edge-on APBs normal to the (001) plane. The elimination of the 
volume-dependent elastic energy mentioned above 
is here possible only for the (100) and (010)-oriented APDs
separated by the (110) or (1$\bar 1$0)-oriented APBs, while in 
the (001)-oriented APDs C and
$\overline {\rm C}$ this elastic energy is always present. Therefore, 
the tweed stage (ii) in these simulations corresponds to both the
predominance of (110)  or (1$\bar 1$0)-oriented APBs separating domains
A or $\overline {\rm A}$ from B or $\overline {\rm B}$
and the decrease of the portion of domains C and $\overline {\rm C}$  in 
the microstructures. In the 3D case each of three posible types of a
polytwin, that without 
(001), (100), or (010)-oriented APDs, can be formed
 in the given part of an alloy stochastically due to the local fluctuations of 
composition [1--7]. It is illustrated, in particular, by 
3D simulation  shown in figure \ref{m4s-3D}, while quasi-2D simulations
describe the formation of only one polytwin type mentioned above.

\begin{figure}
\begin{center}
\epsfig{file=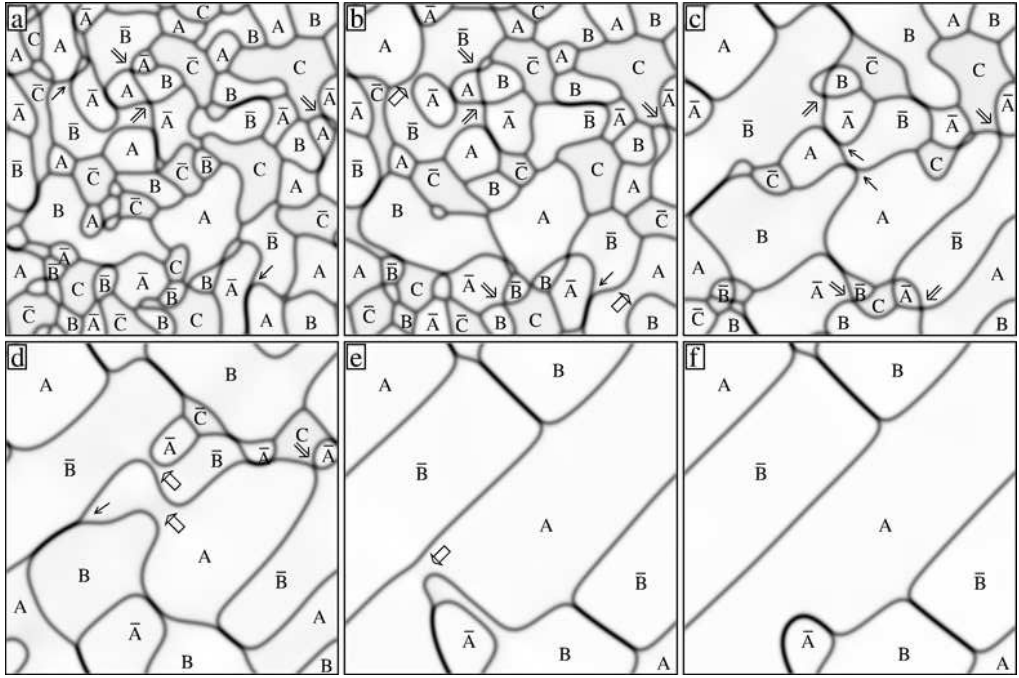,angle=-90,width=0.75\textwidth}
\end{center}
\caption{Temporal evolution of the extended-interaction-range model~5
under the phase transformation A1$\to$L1$_0$ shown in the $\eta^2$-representation
for the simulation box size $V_b=128^2\times 1$ at the maximum tetragonal distortion
parameter $|\varepsilon_m | =0.1$,
$c=0.5$, the reduced temperature  $T'=T/T_c=0.7$,  
 and the following values of the 
reduced time $t'=t\gamma_{nn}$: (a)~10; (b)~20;
(c)~50; (d)~100; (e)~250; and (f)~280.
The grey level linearly varies with 
$\eta_i^2=\eta_{1i}^2+\eta_{2i}^2+\eta_{3i}^2$ between its minimum and
maximum values from completely dark to completely bright. The symbol A,
$\overline {{\rm A}}$, B, $\overline {\rm B}$, C or $\overline {\rm C}$ indicates 
the type of the ordered domain as described in the text.
The thick, the thin and the single arrow indicates the fusion-of-domain process,
the quadruple junction of APDs, and the splitting APB process, 
respectively, discussed in the text.}
\label{m5s_e10}
\end{figure}

\begin{figure}
\begin{center}
\epsfig{file=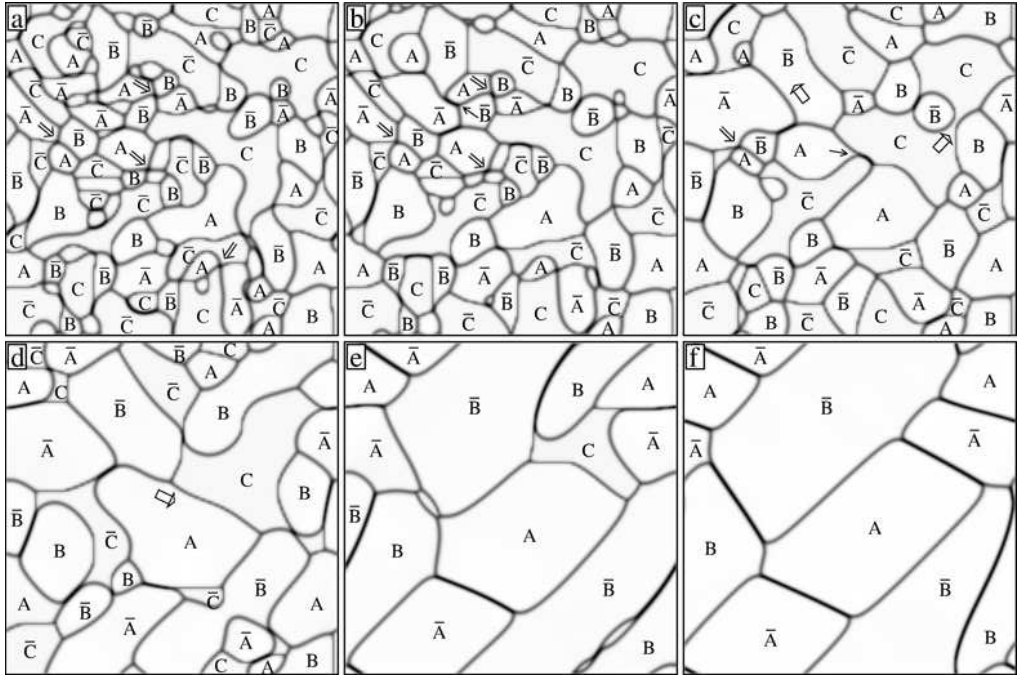,angle=-90,width=0.75\textwidth}
\end{center}
\caption{As  figure \ref{m5s_e10}, but 
for the intermediate-interaction-range model 4 at $T'=0.67$ 
and the following values of $t'$: 
(a)~10; (b)~20; (c)~50; (d)~100; (e)~250;  and (f)~500.}
\label{m4s_e10-1300}
\end{figure}

\begin{figure}
\begin{center}
\epsfig{file=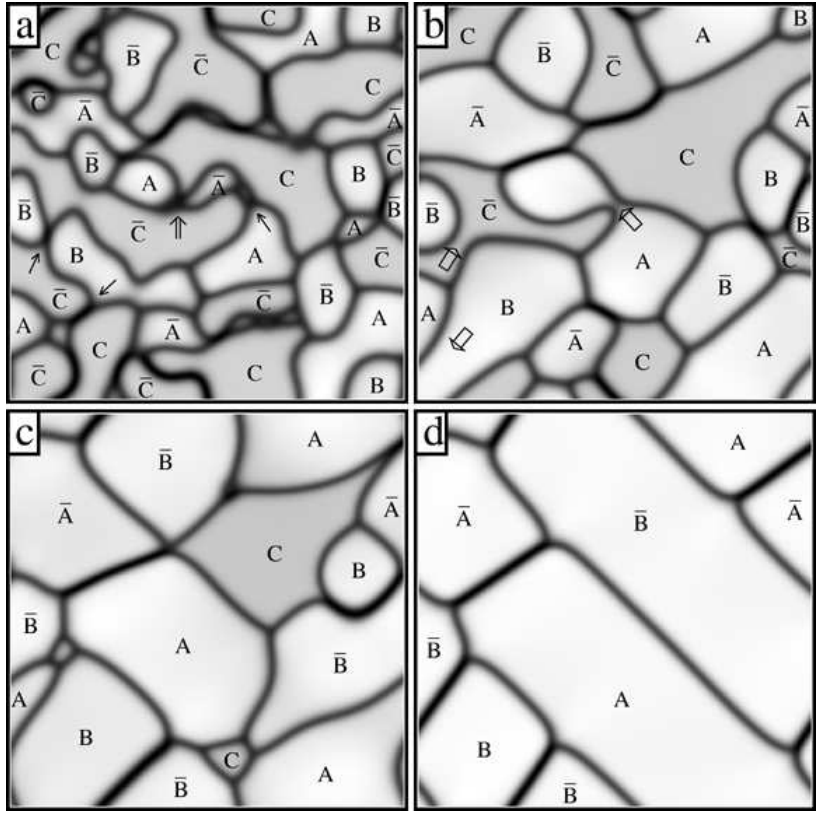,angle=0,width=0.6\textwidth}
\end{center}
\caption{As  figure \ref{m4s_e10-1300}, but 
at $T'=0.92$
and the following values of $t'$: 
(a)~10; (b) 50; (c)~100;   and (d)~200.}
\label{m4s_e10-1800}
\end{figure}

\begin{figure}
\begin{center}
\epsfig{file=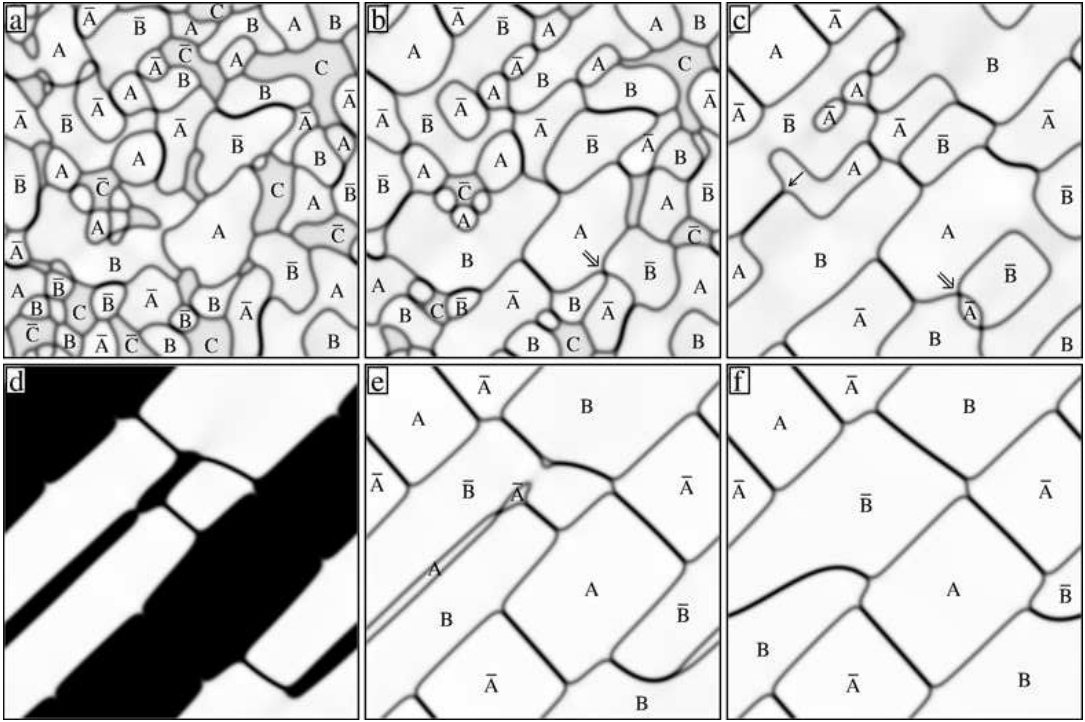,angle=-90,width=0.8\textwidth}
\end{center}
\caption{As  figure \ref{m5s_e10}, but
at $|\varepsilon_m | =0.15$
 and the following values of  $t'$: (a)~10; (b) 20; (c)~50; (d)~150; (e)~172;  and (f)~350.
Frame 2d is shown in the $\eta_2^2$-representation:
the grey level linearly varies with 
$\eta_{2i}^2$ between its minimum and
maximum values from completely dark to completely bright.}
\label{m5s_e15}
\end{figure}

\begin{figure}
\begin{center}
\epsfig{file=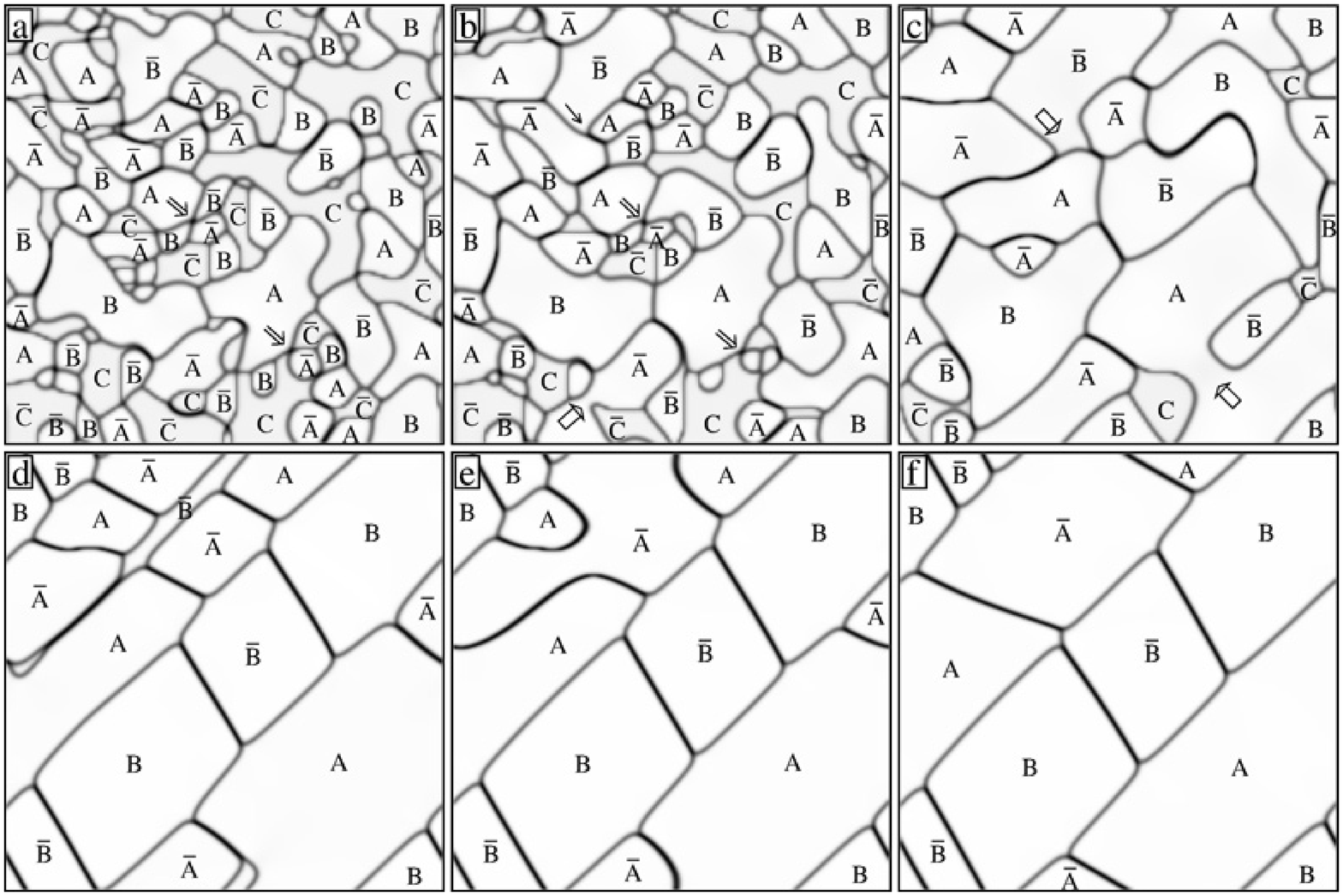,angle=-90,width=0.8\textwidth}
\end{center}
\caption{As  figure \ref{m4s_e10-1300}, but at
$|\varepsilon_m | =0.15$ 
and the following values of $t'$: 
(a)~10; (b)~20; (c)~50; (d)~170; (e)~200;  and (f)~300.}
\label{m4s_e15}
\end{figure}

\begin{figure}
\begin{center}
\epsfig{file=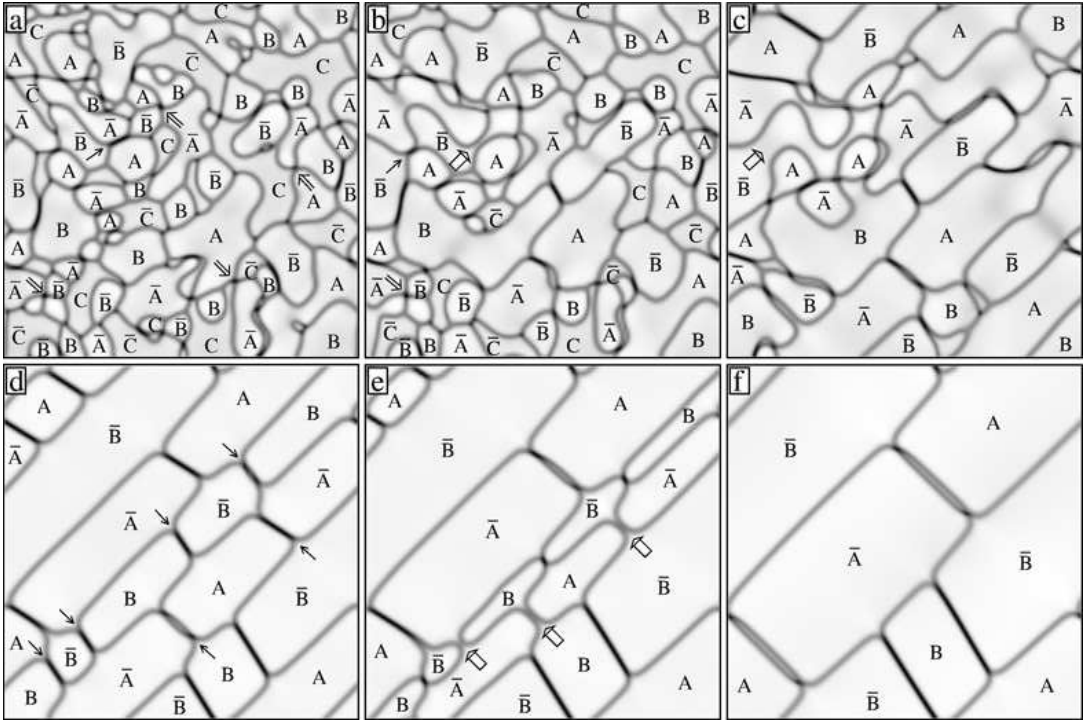,angle=-90,width=0.8\textwidth}
\end{center}
\caption{As  figure \ref{m4s_e15}, but 
at $c=0.44$ 
and the following values of $t'$: 
(a)~10; (b)~20; (c)~50; (d)~400; (e)~750;  and (f)~1100.}
\label{m4ns_e15}
\end{figure}

\begin{figure}
\begin{center}
\epsfig{file=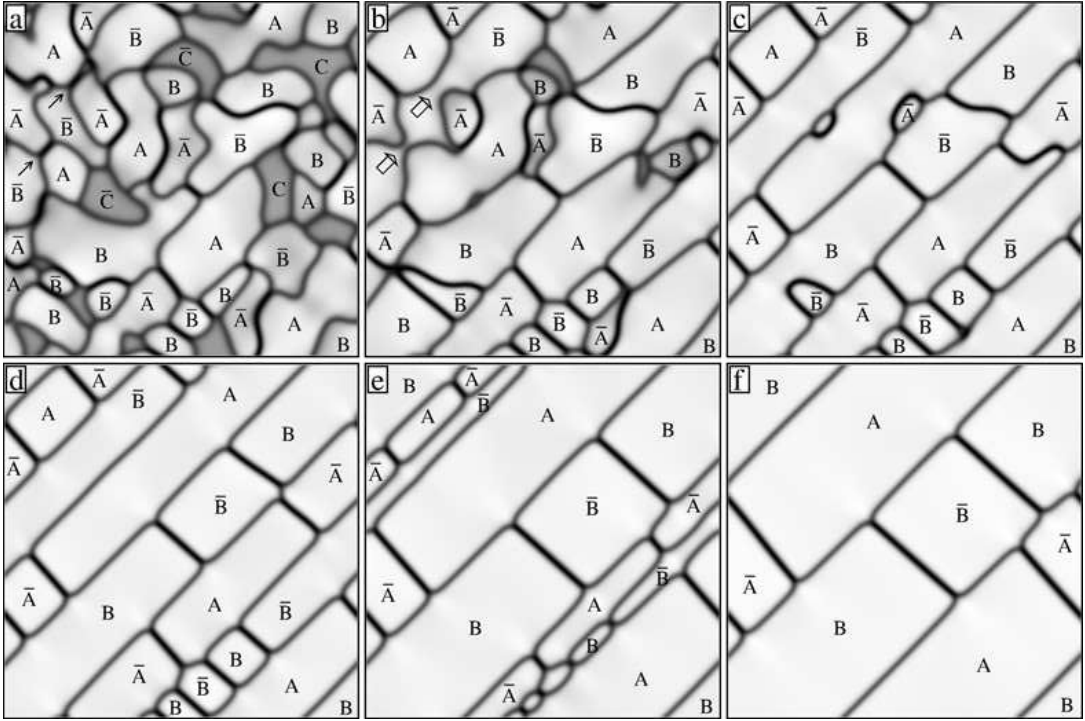,angle=-90,width=0.8\textwidth}
\end{center}
\caption{As  figure \ref{m5s_e10}, but at $|\varepsilon_m | =0.2$, $T'=0.88$ 
and the following values of $t'$: 
(a)~10; (b)~20; (c)~30; (d)~50; (e)~300;  and (f)~400.}
\label{m5s_e20}
\end{figure}

\begin{figure}
\begin{center}
\epsfig{file=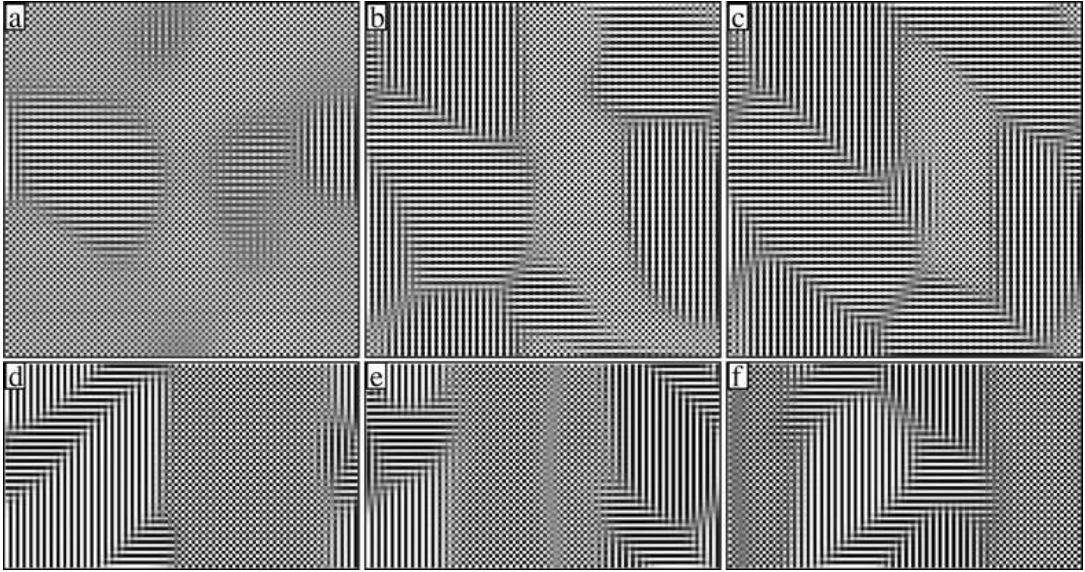,angle=-90,width=0.8\textwidth}
\end{center}
\caption{As  figure \ref{m4s_e10-1800}, but for 3D simulation with $V_b=52^2\times 30$ 
at $|\varepsilon _m|=0.2$ and shown in the `$c$-representation':
the grey level linearly varies with 
$c_i$ between its minimum and
maximum values from completely dark to completely bright.
 The three upper frames correspond to the plane $z=10\,a$ 
and the following values of $t'$: 
(a)~10; (b) 20; and (c)~325. The three lower frames correspond to $t'=325$ and the 
following planes: (d)~$y=0$; (e)~$y=10\,a$;  and (f)~$y=36\,a$.}
\label{m4s-3D}
\end{figure}

The distortion parameter $|\varepsilon_m |=0.1$ for the simulations shown 
in figures \ref{m5s_e10}--\ref{m4s_e10-1800}
was chosen so that the APD size $l_0$ in Eq. (\ref{l_0})  characteristic for 
manifestations of elastic effects has the scale typical for real CoPt-type
alloys. In particular, if we take a conventional assumption that the
APB energy $\sigma$ is proportional to the transition
temperature $T_c$: $\sigma\sim T_cf(T')$ where $f$ is some function of the reduced temperature
$T'=T/T_c$,  then using the relation
$\varepsilon_m =\overline\varepsilon /4\eta_1^2$ and the
parameters $\overline\varepsilon$,
$\eta_1$, and $T_c$ for CoPt and for our models 
mentioned above we find that 
the right-hand side of Eq. (\ref{l_0}) for models 
5 and 4 at $|\varepsilon_m |=0.1$ is close to
that for the CoPt alloy at similar $T'$ values within about ten percent. 
Therefore, the microstructures at both the initial stage (i) and the 
tweed stage (ii) can be reproduced by 
 figures \ref{m5s_e10}--\ref{m4s_e10-1800}
with no significant distortion of scales. Under a furher growth of an APD
its size $l$ becomes comparable with the simulation box size $L$,
and the periodic boundary conditions  begin to significantly 
affect the evolution. Therefore, the later stages of transformation
 can be more adequately simulated if we reduce the characteristic size $l_0$ 
in Eq. (\ref{l_0}) using the  larger values of the parameter $\varepsilon_m$, 
such as  $|\varepsilon_m |=0.15-0.2$ used in the simulations shown 
in figures \ref{m5s_e15}--\ref{m4s-3D}.

Let us first discuss  figures \ref{m5s_e10}--\ref{m4s_e10-1800}  corresponding to
a `realistic' distortion parameter $|\varepsilon_m |=0.1$. The initial stage (i) in these
figures corresponds to frames \ref{m5s_e10}(a)--\ref{m5s_e10}(b), 
\hbox{\ref{m4s_e10-1300}(a)--\ref{m4s_e10-1300}(b)}, and 
\ref{m4s_e10-1800}(a)--\ref{m4s_e10-1800}(b); the tweed stage (ii), to frames
\ref{m5s_e10}(c)--\ref{m5s_e10}(d), \ref{m4s_e10-1300}(c)--\ref{m4s_e10-1300}(e),
and \ref{m4s_e10-1800}(c); and the twin stage (iii),  to frames
\ref{m5s_e10}(e)--\ref{m5s_e10}(f), \ref{m4s_e10-1300}(f),
and \ref{m4s_e10-1800}(d).

The detailed consideration of the initial stage for models 4 and 5 
neglecting the deformational effects  \cite{PV} revealed the following 
features of evolution:

(a) The presence of abundant processes of fusion of in-phase domains which are
 one of  main mechanisms of  domain growth  at this stage.

(b) The presence of peculiar long-living configurations, the
 quadruple junctions of APDs (4-junctions) of the type
A$_1$A$_2$$\overline{\rm A_1}$A$_3$ where A$_2$ and A$_3$ can correspond
to any two of four types of APD different from A$_1$ and $\overline{\rm A_1}$.

(c) The presence of many processes of `splitting' of a shift-APB into two 
flip-APBs which lead to either a fusion of in-phase domains 
mentioned in point (a) ($s\to f$ process) or a formation of a 4-junction 
mentioned in point (b) ($s\to 4j$ process).

Figures \ref{m5s_e10}--\ref{m4s_e10-1800} show that all these microstructural features are
also present when the deformational effects are taken into account, and not only at 
the initial stage (i) but also at the tweed stage (ii). In particular, the beginning and the end of 
an $s\to f$ process (marked by the single and the thick arrow, respectively)
can be followed in 
frames \ref{m5s_e10}(a) and \ref{m5s_e10}(b);
\ref{m5s_e10}(c)  and \ref{m5s_e10}(d) ;  
\ref{m5s_e10}(d)  and \ref{m5s_e10}e; 
\ref{m4s_e10-1300}(b) and \ref{m4s_e10-1300}(c); 
\ref{m4s_e10-1300}(c) and \ref{m4s_e10-1300}(d); and 
\ref{m4s_e10-1800}(a) and \ref{m4s_e10-1800}(b).
The fusion with the disappearance
of an intermediate APD which initially separates two in-phase domains to be fused
\cite{PV} can be followed in the lower right part of frames \ref{m5s_e10}(a) and \ref{m5s_e10}(b)
and in the upper right part of frames \ref{m4s_e10-1300}(b) and \ref{m4s_e10-1300}(c)
(which is marked by a thick arrow in frames \ref{m5s_e10}(b) and \ref{m4s_e10-1300}(c),
respectively).
A number of long-living 4-junctions marked by thin arrows  are seen in frames 
\ref{m5s_e10}(a) --\ref{m5s_e10}(d) , 
\ref{m4s_e10-1300}(a)--\ref{m4s_e10-1300}(c), and 
\ref{m4s_e10-1800}(a). An $s\to 4j$ process can be followed
in the lower right part of frames \ref{m5s_e10}(a)--\ref{m5s_e10}(c) .
The processes and configurations 
(a), (b) and (c) can also be seen in figures 4--7 discussed below.

Frames \ref{m4s_e10-1300}(a)--\ref{m4s_e10-1300}(e) also display
some (100)-oriented and thin conservative APBs. As discussed in  \cite{PV} and below, such   APBs 
are most typical of the short-range-interaction systems where they
have a low surface energy 
(being zero  for  the  stoichiometric nearest-neighbor interaction model) unlike other, non-conservative APBs.
Under an increase of the interaction range, as well as temperature or the deviation
from stoichiometric composition, the anisotropy in the APB surface energy sharply 
decreases \cite{PV}. Therefore, in figure \ref{m4s_e10-1300} (and figure \ref{m4s_e15} below) corresponding to
the intermediate-range-interaction  model 4 
the conservative APBs are few but observable, while for the extended-range-interaction model 5 
in   figure \ref{m5s_e10}, as well as at elevated $T$ or significant `non-stoichiometry'
  $\delta c=(0.5-c)$
in figures \ref{m4s_e10-1800} or  \ref{m4ns_e15}  for model 4, such APBs are absent entirely.

Comparison of figures \ref{m4s_e10-1300} and \ref{m4s_e10-1800} illustrates the sharp dependence
of microstructural evolution on the transformation temperature $T$.  Under elevating this temperature
to values near the critical temperature $T_c$: $(T_c-T)\lesssim 0.1\,T_c$,
both flip and shift-APBs notably thicken, the anisotropy in their surface energy falls off,  while
the characteristic size of initial APDs (formed after a rapid quench A1$\to$L1$_0$) increases.
The latter is related to an increase   at $T\to T_c$ of the characteristic 
wavelength for the ordering instability
which is due to the narrowing of the interval of effective wavenumbers
${\bf q}={\bf k}-{\bf k}_s$ near the superstructure vector ${\bf k}_s$ for which 
the ordering concentration waves are unstable at $T<T_c$.

Frames \ref{m5s_e10}(c)--\ref{m5s_e10}(d), \ref{m4s_e10-1300}(c)--\ref{m4s_e10-1300}(e),
and \ref{m4s_e10-1800}(c) show evolution  at the tweed stage. They  illustrate, in particular, kinetics of 
the (110)-type alignment of APBs  between APDs A or $\overline {\rm A}$ 
and B or $\overline {\rm B}$ at this stage, as well as 
 a "dying out" of  (100)-oriented APDs C and $\overline{\rm C}$. These frames 
also show  that in the simulation with a realistic distortion parameter
 $|\varepsilon_m|=0.1$ 
(fitted to the structural data for CoPt)
the  APD size $l_0$ (\ref{l_0}) characteristic 
of the tweed stage is about $(20-40)\,a$. It agrees with the order of magnitude 
of this size observed in the CoPt-type alloys FePt and FePd \cite{Zhang-91,Zhang-92,Yanar}.

As mentioned, the final, twin stage of the transformation can be more adequately simulated 
with the larger values of parameter $|\varepsilon_m|$ which are employed 
in the simulations shown in figures  \ref{m5s_e15}--\ref{m4s-3D}. Before discussing
 these figures we note some typical configurations
 observed in experimental studies of transient twinned microstructures \cite{Zhang-91,Zhang-92,Yanar}
seen, for example,
in figures 5, 6, 9 and 2 in Refs. \cite{Zhang-91} \cite{Zhang-92}, \cite{Yanar} and \cite{CWK}, respectively:

(1) semi-loop-like shift-APBs adjacent to the twin band boundaries;

(2) `S-shaped'  shift-APBs stretching across the twin band; and

(3) short  and narrow twin bands (for brevity to be called  `microtwins') which lie within the larger twin bands
and usually have one or two shift-APBs near their edges.

Comparing our results with experiments
one should consider that  due to the limited size of the simulation box
the twin band width $d$ in our simulations
 has the same order of magnitude as the 
APD size $l_0$ (\ref{l_0}) characteristic of the tweed stage, while 
in experiments $d$ usually much exceeds $l_0$ \cite{Zhang-91,Zhang-92,Yanar}.
Therefore, the distribution of shift-APBs within twin bands 
in our simulation is usually much more close to  equilibrium than in experiments.
In spite of this difference,  the simulations reproduce 
all characteristic  transient configurations (1)--(3)  and elucidate the  mechanisms
of their formation. In particular, both the semi-loop and the S-shaped  shift-APBs
are  formed from regular-shaped approximately quadrangular APDs
(characteristic of the beginning of the twin stage) due to the disappearance of
adjacent APDs which are `wrongly-oriented' with respect to the given twin band.
The formation of semi-loop configurations is  illustrated by frames 
 \ref{m5s_e10}(d)--\ref{m5s_e10}(f),
 \ref{m4s_e15}(d)--\ref{m4s_e15}(e), and  \ref{m5s_e20}(b)--\ref{m5s_e20}(c);
while the formation of S-shaped APBs can be seen in
frames  \ref{m4s_e10-1300}(d)--\ref{m4s_e10-1300}(f),  \ref{m5s_e15}(d)--\ref{m5s_e15}(f),
 \ref{m4s_e15}(d)--\ref{m4s_e15}(e), and  \ref{m5s_e20}(b)--\ref{m5s_e20}(c).
The formation and evolution of microtwins is illustrated by frames 
\ref{m5s_e15}(c)--\ref{m5s_e15}(d) and  \ref{m4s_e15}(c)--\ref{m4s_e15}(d).
These frames show that the microtwin is actually a small and narrow APD
for which deformational effects are strong enough to align its flip-APBs
along the (110)-type directions. However, the standard mechanism of coarsening
via the growth of larger APDs at the expence of  smaller ones
leads to the shrinking and eventually to the disappearance of a microtwin which
is usually accompanied by the formation of S-shaped and/or semi-loop
shift-APBs. The latter  is illustrated by frames  \ref{m5s_e15}(d)--\ref{m5s_e15}(f)
 and  \ref{m4s_e15}(d)--\ref{m4s_e15}(e). Let us also note that the microtwin configuration
shown in frame  \ref{m5s_e15}(d) is strikingly similar to that seen in the central part
of experimental figure 2 in Ref. \cite{CWK}.

Let us now discuss the final, `nearly equilibrium'  microstructures 
shown in last frames of figures \ref{m5s_e10}--\ref{m4s-3D}.
A characteristic  feature of these microstructures is a peculiar  alignment of shift-APBs: 
within a (100)-oriented twin band in a (110)-type
polytwin the APBs tend to align normally to 
some direction ${\bf n}=(\cos\alpha, \sin\alpha, 0)$ characterized by a  `tilting'
angle $\alpha$ between the band orientation and the APB plane. Figures
 \ref{m5s_e10}--\ref{m4s-3D} show that this tilting angle is not very sensitive to
the variations of temperature or concentration but it sharply depends on the 
interaction type, particularly on the interaction range.
For the extended-range-interaction model 5 this angle is close to $\pi/4$
(slightly exceeding this value), while for  the  intermediate-range-interaction model~4
it is notably less than $\pi/4$. A similar alignment of APBs for the short-range interaction systems
 is illustrated by figures  \ref{m1s_e15}--\ref{m2_e10} where the tilting angle is
close to zero. 

A phenomenological theory of this interaction-dependent  tilting of APBs within 
nearly-equilibrium twin bands has been suggested in \cite{Vaks-01}. The tilting is
explained  by the competition 
 between the anisotropy of the APB surface tension $\sigma$
and a tendency to minimize the total APB area within the given twin band which
corresponds to $\alpha =\pi/4$. For the alloy systems with both the intermediate and the short interaction range
the surface tension $\sigma (\alpha)$ is minimal at $\alpha =0$ \cite{Vaks-01,PV}. Thus 
for such alloy systems the tilting angle is less than
$\pi/4$, and it decreases with the decrease of the interaction range. For the extended-range-interaction
systems the anisotropy of the APB surface tension is weak \cite{BDPSV,PV}, and so the tilting angle is close to $\pi/4$.
Therefore, the comparison of
experimental tilting angles with theoretical calculations \cite{Vaks-01}
can provide both qualitative and quantitative
information on the effective chemical interactions in an alloy.

The alignment of shift-APBs discussed above seems  to be clearly seen in the 
experimental microstructure for CoPt  shown in figure 5 of Ref. \cite{Leroux} where the tilting angle
 is notably less than  $\pi/4$. It can 
indicate that the effective  interactions in  CoPt have an intermediate interaction range. This
 agrees with the usual estimates of these interactions for transition metal alloys,
see e.g. \cite{Zunger,Chassagne,Turchi}.

Comparison of figures \ref{m5s_e15} and  \ref{m5s_e20} illustrates the influence of
temperature $T$ on the evolution. Under elevating $T$ we again observe a thickening of APBs,
as well as a coarsening of  initial APDs. 
Frames  \ref{m5s_e20}(d)--\ref{m5s_e20}(f) also illustrate a process of `transverse coarsening'
of twin bands via a shrinkage and disappearance of some microtwinned bands. Such 
transverse coarsening appears to be seen in a number of experimental microstructures,
for example, in figure 6 in \cite{Zhang-92} or figure 2 in \cite{CWK}. Frames 
\ref{m5s_e20}(d)--\ref{m5s_e20}(f) show that the thermodynamic driving force for 
 such transverse coarsening is mainly the gain in  the surface energy of shift-APBs due  
the decrease of their total area under this process.

Figures \ref{m4s_e15} and  \ref{m4ns_e15} illustrate the concentration dependence of
the  evolution. The non-stoichiometry $\delta c =(0.5-c)$ affects the
evolution similarly to temperature $T$: under an increase of both 
$\delta c$ and $T$ all APBs thicken, while shift-APBs become less stable with respect to 
flip-APBs \cite{PV}. The latter leads to an enhancement of processes of splitting of shift-APBs
as well as of the transverse coarsening mentioned above; it is illustrated by frames 
 \ref{m4ns_e15}(b)-- \ref{m4ns_e15}(f).

 Some results of a 3D simulation with $V_b=52^2\times 30$ are presented in figure \ref{m4s-3D}.
 In this figure  we employ the $c$-representation (described in the caption) in which
the regions containing the vertical or horizontal lines (that is, the
 vertical or horizontal crystal planes filled by A atoms) correspond to the APDs 
with the (100) or (010)-oriented tetragonal axis, respectively, while the checkered regions
correspond to the APDs with the tetragonal axis normal to the plane of figure. 
This simulation aimed mainly to complement 2D simulations
with an illustration of geometrical features of 3D microstructures. Figure  \ref{m4s-3D}
 illustrates, in particular, a stochastic formation of different polytwin sets with 
three possible types of orientation mentioned above. A limited size of the simulation box
prevent us from a detailed consideration of  evolution with this 3D simulation.
Therefore, below we discuss only the problem of a 3D orientation of tilted shift-APBs in 
final, `nearly-equilibrium' microstructures. 

Let us consider a  (100)-oriented twin band
in the form of a plate of  height $h$,  length $l$, and width $d$  in the direction (001), 
(110), and  (1$\bar 1$0), respectively, with $d\lesssim h\ll l$ (which is a typical form of 
twin bands observed in TEM experiments 
\cite{Leroux,Zhang-91,Zhang-92,Yanar,Tanaka,Oshima,Syutkina}).
The equilibrium orientation of a plane shift-APB in this band corresponds to the minimum
of its energy $E_s=\sigma S$ where $S$ is the APB area and 
$\sigma$ is the surface tension determined mainly by the 
angle $\alpha$ between 
the APB orientation ${\bf n}=(\sin\alpha, \cos\alpha\cos\varphi, \cos\alpha\sin\varphi)$
and the band orientation ${\bf n}_0=(100)$ \cite{Vaks-01}.
For the `needle-shaped' twin band under consideration 
the upper and the lower boundary of a shift-APB usually lies
at the top and the bottom edge of this band, respectively.
Minimization of energy $E_s$ in this case yields: $\varphi =0$, 
i. e. the APB is normal to the (001) plane,
and its orientation ${\bf n}=(\sin\alpha, \cos\alpha, 0)$ 
is determined by the interaction-dependent tilting angle
$\alpha$ defined in \cite{Vaks-01}. This conclusion  seems to be supported 
by the present 3D simulation:
the lower and the upper tilted shift-APB within the (010)-oriented twin band 
below the main diagonal of frame \ref{m4s-3D}(c)
corresponds to the grey line stretching across the checkered region in 
frame \ref{m4s-3D}(e) and \ref{m4s-3D}(f), respectively, 
and both these lines are approximately normal to the (001) plane.

\section{Kinetic features of A1$\to$L1$_0$ transformations in 
the short-range interaction systems}

As mentioned,  transient microstructures under L1$_0$ ordering 
for  the short-range interaction systems
include many conservative APBs.
Such APBs are virtually immobile, and so the evolution is realized
via motion of other, non-conservative  APBs which results in a number 
of peculiar kinetic features \cite{BDPSV,PV}. The initial stage of  
the A1$\to$L1$_0$  transformation for the short-range interaction systems
was discussed in detail in Ref. \cite{PV}. In this section we consider 
 the tweed and  twin stages of such transformations and note the 
 differences with the  case of systems 
with the larger interaction range.
  
\begin{figure}
\begin{center}
\epsfig{file=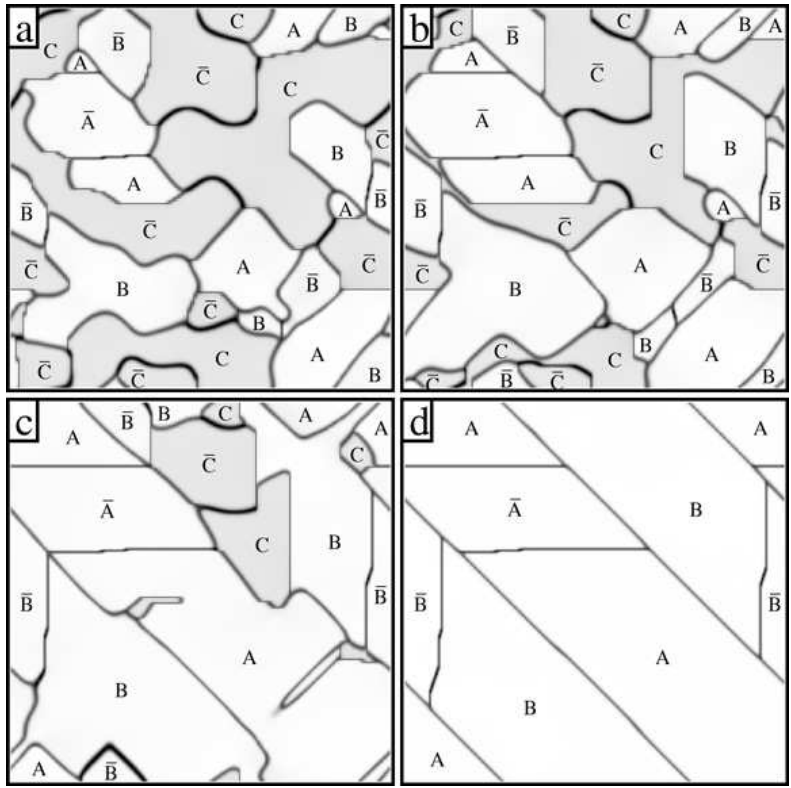,angle=0,width=0.58\textwidth}
\end{center}
\caption{As figure \ref{m5s_e10} but for the short-range-interaction model 1
at $|\varepsilon_m|=0.15$, $c=0.5$,
$T'=0.9$ and the following values of $t'$: (a) 30;  (b) 40; (c) 60; and (d) 120.}
\label{m1s_e15}
\end{figure}

\begin{figure}
\begin{center}
\epsfig{file=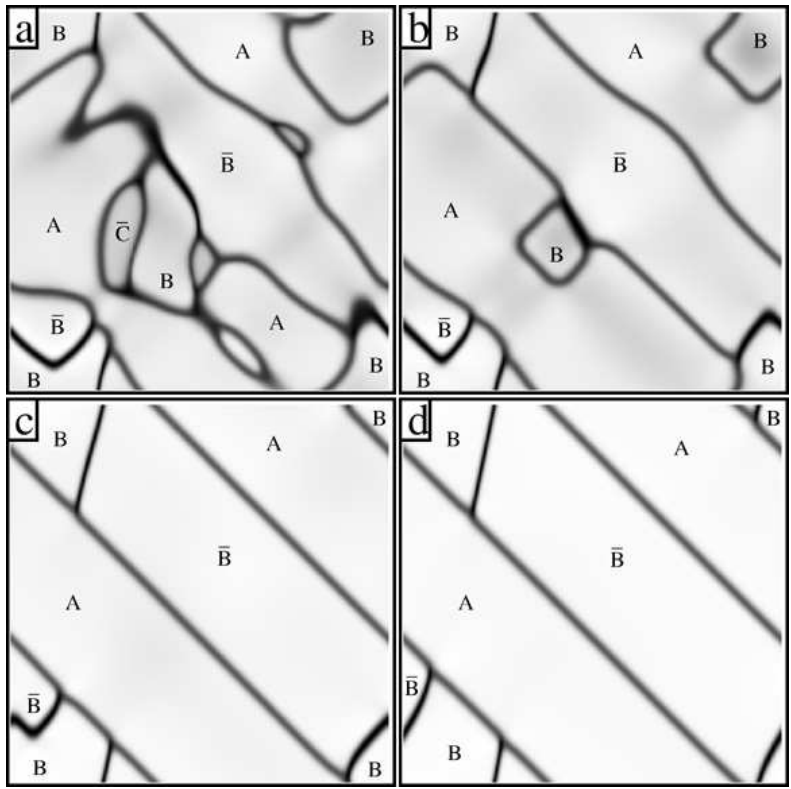,angle=0,width=0.58\textwidth}
\end{center}
\caption{As figure \ref{m1s_e15} but at $c=0.44$
 and the following values of $t'$: (a)~110;   (b)~140;  (c)~200;  and (d)~350.}
\label{m1ns_e15}
\end{figure}

\begin{figure}
\begin{center}
\epsfig{file=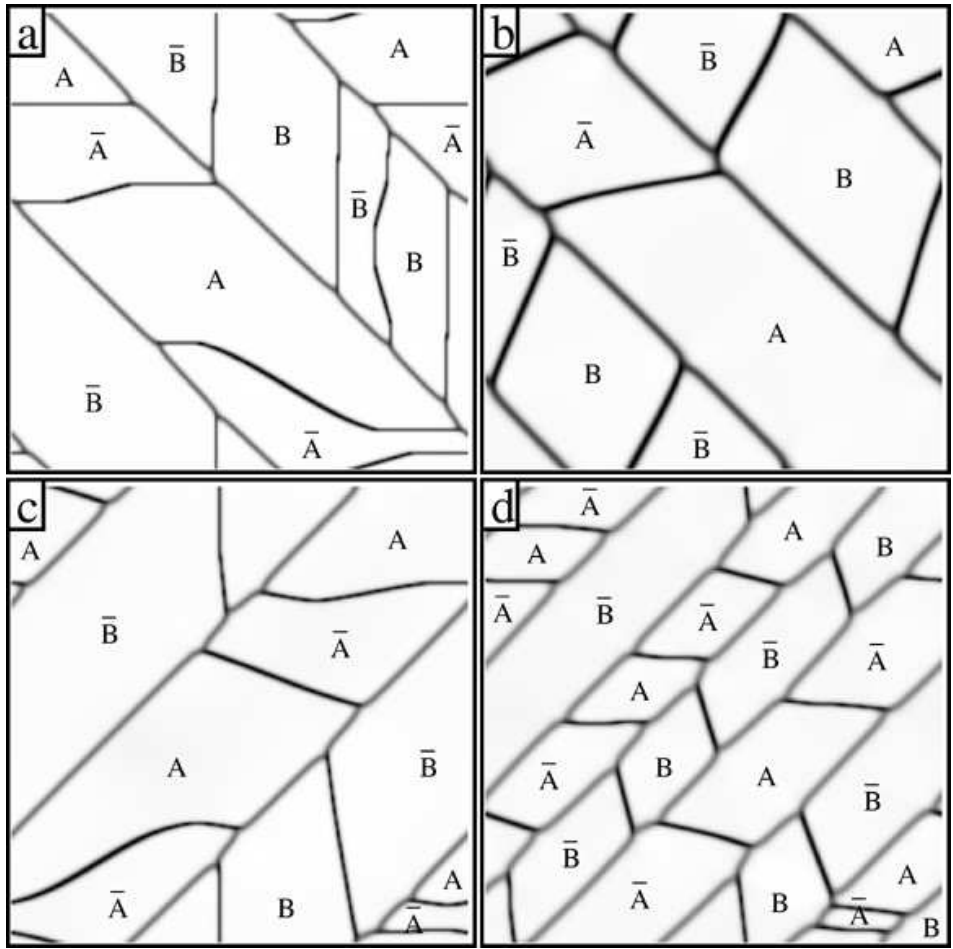,angle=0,width=0.7\textwidth}
\end{center}
\caption{As figure \ref{m5s_e10} but for model $2$
at  $|\varepsilon_m|=0.1$  and the following values of  $c$, $T'$ and $t'$: 
(a) $c=0.5$, $T'=0.77$, $t'=350$;
(b) $c=0.5$, $T'=0.95$, $t'=300$;
(c) $c=0.46$, $T'=0.77$, $t'=350$;
and (d) $c=0.44$, $T'=0.77$, $t'=300$.}
\label{m2_e10}
\end{figure}

Some results of our simulations for the short-range-interaction systems  are presented in figures
\ref{m1s_e15}--\ref{m2_e10}. 
In these simulations we used sufficiently high temperatures $T'\gtrsim 0.9-0.8$ to
accelerate evolution to final, `nearly-equilibrium' configurations as the presence of
conservative APBs slowes down (or even `freezes') this evolution, particularly at low $T'$.

Figure  \ref{m1s_e15} illustrates the evolution for model 1;  as discussed in \cite{BDPSV}, 
this model seems to correspond to the Cu--Au-type alloys.
A distinctive feature of microstructures shown in figure \ref{m1s_e15} 
is a predominance of the above-mentioned conservative APBs with the (100)-type orientation.
Frames \ref{m1s_e15}(a)--\ref{m1s_e15}(c) show both the
conservative shift-APBs
\hbox{(c-shift APBs)} and the conservative flip-APBs
\hbox{(c-flip APBs)} also illustrating their orientational properties \cite{PV}. 
For quasi-2D  microstructures with edge-on APBs shown in  figure \ref{m1s_e15},
c-shift APBs separating APDs A and \hbox{$\overline{{\rm A}}$} (c-APBs \hbox{A--$\overline{{\rm A}}$)}
are horizontal; c-APBs  \hbox{B--$\overline{\rm B}$} are vertical; and c-APBs  
\hbox{C--$\overline{\rm C}$} can be both horizontal and vertical;
 c-flip APBs \hbox{(A or $\overline{\rm A}$)}--\hbox{(C or $\overline{\rm C}$)}
(which separate  APDs A or $\overline{\rm A}$ from C or $\overline{\rm C}$)
are horizontal; c-APBs \hbox{(B or $\overline{\rm B}$)}--\hbox{(C or $\overline{\rm C}$)} are vertical;
and c-APBs \hbox{(A~or~$\overline{\rm A}$)}--\hbox{(B~or~ $\overline{\rm B}$)} should lie
in the plane of figure and thus they are not seen in figure \ref{m1s_e15}.
Figure \ref{m1s_e15} also shows that the conservative APBs are notably thinner than 
non-conservative ones, particularly so for c-flip APBs.

Frames \ref{m1s_e15}(a)--\ref{m1s_e15}(c) show that at first stages of evolution 
the portion of conservative APBs with respect to non-conservative ones increases, 
due to the lower surface energy of the c-APBs. Later on, with the beginning of the tweed stage, the 
deformational effects become important leading to a dying out of both APDs C and $\overline{\rm C}$ 
and their \hbox{c-flip APBs}. However, the conservative shift-APBs within twin bands 
survive, and  in the final frame  \ref{m1s_e15}d they are mostly `step-like' 
consisting of (100)-type oriented conservative segments and small
non-conservative ledges. 
 These step-like APBs  can be viewed as  a `facetted'  
version of  tilted APBs discussed above and seen in  figures \ref{m5s_e10}--\ref{m4s-3D}. 
Such step-like APBs were observed under the L1$_0$ ordering of 
CuAu and some CuAu-based alloys \cite{Syutkina}, and they are also
similar to those 
observed under the L1$_2$ ordering in both simulations \cite{BDPSV} and 
experiments for the Cu$_3$Au alloy  \cite{Potez}.

As it was repeatedly noted in Ref. \cite{PV} and above, 
 an increase of  non-stoichiometry  $\delta c= (0.5-c)$ or temperature $T$
leads to a sharp decrease of both the anisotropy of the APB energy and
 the energy preference of conservative APBs with respect to
non-conservative ones. Therefore, under an increase of $\delta c$ or $T$ 
the portion of conservative APBs in transient microstructures falls off,
and at sufficiently high $\delta c$ or $T$ such APBs are not formed under the transformation  at all.
 It results in drastic microstructural changes of evolution,
including  sharp, phase-transition-like changes in morphology of 
aligned shift-APBs within twin bands,  from  the  `faceting' to the
`tilting'. This is illustrated by figure \ref{m1ns_e15} which shows the
evolution of model 1 at a significant non-stoichiometry $\delta c =0.06$,
and this evolution is qualitatively different with that for a stoichiometric alloy shown in figure  \ref{m1s_e15}. 

Figure \ref{m2_e10} illustrates the  transition from the `facetted' to the `tilted' morphology of
shift-APBs within nearly-equilibrium twin bands under variations of $T$ or $\delta c$ for model 2.
 An examination of intermediate
stages of transformations illustrated by this figure  shows that 
the morphological changes are  realized via some local bends 
of facetted APBs. It is also  illustrated by a comparison of frames  \ref{m2_e10}(a), \ref{m2_e10}(c)
and \ref{m2_e10}(d) with each other.
Therefore, the `morphological phase transition' mentioned above is 
actually smeared over some interval of temperature or concentration.
However, frames  \ref{m2_e10}(a)--\ref{m2_e10}(d) show that the `intervals
of smearing' of such transitions  can be relatively narrow.

\section{Conclusion}

Let us summarize the main results of this work.
The earlier-described   master equation approach  \cite{Vaks-96,BV-98}
is used to study the microstructural evolution  under L1$_0$-type
orderings in alloys, including the formation of 
 twinned structures due to the spontaneous tetragonal deformation 
inherent to such orderings. To this end we first derive a microscopical model
for the effective interatomic deformational interaction 
 which arise due to the so-called  Kanzaki forces
describing interaction of lattice deformations with site occupations.
This model  generalizes an analogous model
of Khachaturyan for dilute alloys 
\cite{Khach-book} to the physically interesting case of concentrated alloys. 
We take into account the non-pairwise contribution to
Kanzaki forces, and the resulting effective interaction $H_d$ is
non-pairwise, too, unlike the case of dilute alloys.  This effective
interaction describes, in particular, the
lattice symmetry change  effects under phase transformations, such as the tetragonal
distortion mentioned above. Assuming the non-pairwise Kanzaki forces
 to be short-ranged, we can express the deformational interaction $H_d$ in terms of two 
microscopical parameters which  can be estimated from the experimental data about
the lattice distortions under phase transformations. We present these estimates
for alloys Co--Pt for which such structural   data are available \cite{Leroux-88}. 

Then we employ the kinetic cluster field method \cite{BDPSV,PV}
to simulate A1$\to$L1$_0$ transformation 
after a quench of an alloy from the disordered 
A1 phase to the single-phase L1$_0$ field of the phase diagram in 
the presence of deformational interaction $H_d$.
We consider five  alloy models
with different types of chemical interaction, from the short-range-interaction model 1
to the extended-range-interaction model 5, at different temperatures $T$, concentrations $c$, 
and spontaneous tetragonal distortions $\overline\varepsilon$. 
We use both 2D and 3D
simulations, and all significant features of microstructural evolution 
in both types of simulation were found to be similar.
 
The evolution under A1$\to$L1$_0$  transition can be divided into
three stages, in accordance with an increasing importance of the deformational
interaction $H_d$: the `initial', `tweed' and `twin' stage. For the initial
stage (discussed in detail previously \cite{PV}), the deformational effects are insignificant.
For the tweed stage, the effects of $H_d$ become comparable with those of
chemical interaction $H_c$ and lead to the formation of specific microstructures
discussed  in section 4.
For the  final, twin stage the tetragonal distortion
of L1$_0$-ordered antiphase domains (APDs) becomes
the main factor of the evolution and leads to the
formation of (110)-type oriented twin bands.
 Each band includes only two types of APD with the same tetragonal axis, and  
these axes in the adjacent bands  are `twin' related, \hbox{i. e.}  have the 
alternate (100) and (010) orientations for the given set of (110)-type oriented bands.

 The microstructural evolution 
strongly depends on the interaction type, particularly on the interaction range $R_{int}$.
For the systems with an extended or intermediate $R_{int}$ 
at both the initial and the tweed stage we observe the following
features (mentioned previously  \cite{PV} for the initial stage):
(a) abundant processes of fusion of in-phase domains;
(b) a great number of peculiar long-living configurations, the
 quadruple junctions of APDs described in section 4; and 
(c)  numerous processes of `splitting' of an antiphase boundary separating the APDs with the same
tetragonal axis (`shift-APB') into two APBs separating the APDs with perpendicular
tetragonal axis (flip-APBs).
 The simulations also illustrate a sharp temperature dependence of the evolution,
in particular, a notable increase of both the width of APBs and the characteristic size of initial APDs 
 under elevating $T$. 
The deviation from stoichiometry  affects the
evolution similarly to temperature: under an increase of both 
non-stoichiometry $\delta c =(0.5-c)$ and $T$ all APBs thicken, while shift-APBs become less stable with respect to 
flip-APBs.

For the twin stage, our simulations reveal the following typical features of  transient 
microstructures:
(1) semi-loop-like shift-APBs adjacent to the twin band boundaries;
(2) `S-shaped'  shift-APBs stretching across the twin band; 
(3) short  and narrow twin bands (`microtwins') lying within the larger twin bands; and
(4) processes of   `transverse coarsening'
of twinned structures via a shrinkage and disappearance of some microtwins. 
 All these features agree with experimental observations \cite{Zhang-91,Zhang-92,Yanar}.
For the final, nearly-equilibrium twin bands the simulations
demonstrate a peculiar  alignment of shift-APBs
with a certain tilting angle between the band orientation and the APB plane,
and this tilting angle sharply depends on
the interaction type, particularly on the interaction range $R_{int}$.
 Such alignment of APBs seems to be  observed  in the CoPt alloy \cite{Leroux},
 and a comparison of
experimental tilting angles with theoretical calculations \cite{Vaks-01}
can provide information about the effective  interactions in an alloy.

A distinctive feature of  evolution 
for the short-range-interaction
systems  is the presence of many  conservative APBs with the (100)-type orientation.
 The conservative flip-APBs disappear in the course of the evolution,
but the conservative shift-APBs survive and are present in the final twinned 
microstructures. Such `nearly equilibrium'  shift-APBs are mostly 'step-like' 
consisting of (100)-type oriented conservative segments and small
non-conservative ledges, which can be viewed as  a `facetted'  
version of tilted APBs mentioned above.
This (100)-type alignment of shift-APBs within twin bands 
 seems to agree with available experimental observations for the CuAu alloy
\cite{Syutkina} for which chemical interactions are supposed  
to be short-ranged \cite{Potez,BDPSV}.
 
Under an increase of  non-stoichiometry  $\delta c$ or temperature $T$
 the energy preference of conservative APBs with respect to
non-conservative ones decreases, and 
the portion of conservative APBs in the microstructures falls off.
 It results in drastic microstructural changes,
including  sharp, phase-transition-like changes in morphology of 
aligned shift-APBs within twin bands,  from  their  `faceting' to the
`tilting'.  Such `morphological phase transitions' are
actually smeared over some intervals of temperature or concentration,
but the simulations show that the intervals of smearing  can be narrow.

Finally, let us make a general remark about kinetics of
 multivariant orderings in alloys, such as the  L1$_2$, L1$_0$ and D0$_3$ orderings discussed in Refs. 
\cite{BDPSV,PV,BSV}
and in this work.
It is  known that the thermodynamic behavior of different systems
under various phase transitions reveals features of universality and 
insensitivity to the microscopical details of structure,
 particularly in the critical region near thermodynamic instability points. 
The results of this and other studies of  multivariant orderings  
show that such universality 
does not seem to hold for their phase transformation kinetics,
at least outside the critical region (which for 
such orderings  is usually
either quite narrow or absent at all). The microstructural evolution 
 reveals a great variety of peculiar features,
the detailed form of which sharply depends on the type of 
interatomic  interaction, the type of the crystal structure and ordering, the degree of 
non-stoichiometry,   and other `non-universal' characteristics.

\section*{Acknowledgments}

The authors are much indebted to V.~Yu.~Dobretsov for the help in this work;
to N.~N.~Syutkin and V.~I.~Syutkina, for the valuable information about details
of experiments \cite{Syutkina}; and to Georges Martin, for numerous stimulating
discussions. The work was supported  by the Russian Fund of Basic Research under
Grants No 00-02-17692 and 00-15-96709.

\end{document}